\documentclass[11pt,aps,showpacs,nofootinbib,longbibliography]{revtex4-2}

\pdfoutput=1

\usepackage{ulem}

\usepackage{comment}
\usepackage{hyperref}
\usepackage{graphicx}
\usepackage{amsmath,amssymb,color}
\usepackage{grffile}
\usepackage{ulem}
\usepackage{graphicx}
\usepackage{subfig}

\newcommand{\beq}{\begin{equation}}
\newcommand{\eeq}{\end{equation}}

\newcommand{\beqn}{\begin{eqnarray}}
\newcommand{\eeqn}{\end{eqnarray}}

\newcommand{\jo}[1]{{\color{black}#1}}

\begin{document}

\title{
 \jo{Cosmic Birefringence from the Axiverse}
}

\author{
Silvia Gasparotto$^{1,2}$ and Evangelos I. Sfakianakis$^{2,3}$
}

\affiliation{
$^1$Grup de F\'{i}sica Te\`{o}rica, Departament de F\'{i}sica, Universitat Aut\`{o}noma de Barcelona, 08193 Bellaterra (Barcelona), Spain
\\
 $^2$Institut de F\'isica d'Altes Energies (IFAE), The Barcelona Institute of
Science and Technology (BIST), Campus UAB, 08193 Bellaterra, Barcelona
\\
$^3$Department of Physics, Case Western Reserve University,
10900 Euclid Avenue, Cleveland, OH 44106, USA
}\email{sgasparotto@ifae.es, esfakianakis@ifae.es}


\begin{abstract}

{We revisit the evidence for CMB birefringence in the context of a rich Axiverse.  Using probability density functions (PDFs) for various axion parameters,
such as the mass and axion decay constant, we construct the PDF for the
cosmic birefringence angle and investigate its properties. 
By relating the observed value of the birefringence angle to the mean or standard deviation
of the constructed PDF, we  constrain  the shape of the input PDFs, providing insights
into the statistical distribution of the Axiverse. We focus on three different types of axion
potentials: cosine, quadratic, and asymptotically linear axion monodromy.
Our
analysis showcases the potential of cosmic birefringence in constraining the distribution
of axion parameters and uncovering possible correlations among them. 
We additionally offer
predictions for “birefringence tomography,” anticipating future measurements of birefringence from
lower multipoles, and show how it can be used to rule out simpler versions of the Axiverse. Our findings contribute to the
ongoing exploration of the Axiverse and its implications for cosmic birefringence.
}

\end{abstract}

\maketitle

\newpage 

\tableofcontents

\newpage
\section{Introduction}

Several analyses of the CMB data have found an intriguing signal of \textit{Cosmic Birefringence} \cite{Minami:2020odp, Diego-Palazuelos:2022dsq, Eskilt:2022cff,Eskilt:2023ndm}, \jo{which measures the \textit{angle of rotation} of the CMB radiation}, with the most recent value being $\beta=0.342^{+0.094º}_{-0.091º}$, excluding the null result by more than 3$\sigma$. This angle \jo{ refers to the static and} isotropic rotation of the polarization of photons emitted at the ``last scattering" (LS) surface which shows up in CMB experiments, specifically in the EB cross-correlation signal $C^{EB}_l\propto\sin(4\beta)(C_l^{EE,\text{CMB}}-C_l^{BB,\text{CMB}})$ (see Ref.~\cite{Komatsu:2022nvu} for a recent review). This measurement has been notoriously complicated because the birefringence angle $\beta$ is degenerate with the miscalibration of the polarimeters, but the new technique developed in Refs.~\cite{Minami:2019ruj, Minami:2020fin} made it possible to distinguish the two and achieve better precision. The measurement \jo{appears} robust under instrumental systematics, but potential EB contribution from galactic dust currently represents the biggest limit\jo{ation} of the analysis \cite{Diego-Palazuelos:2022cnh}. 

The birefringence signal finds an elegant explanation in the context of axion physics where a light scalar field $\phi$, which can be also tied to dark matter or dark energy, couples to the electromagnetic field strength $F^{\mu\nu}$ via the Chern-Simons interaction $\phi F \Tilde{F}$. This additional term in the Lagrangian breaks the parity of the electromagnetic sector and changes the dispersion relation between the left and right-handed photons. This effectively makes the Universe, filled with such scalar field $\phi$, behave as a birefringent material {with a different refraction index for the right and left-handed polarization. Thus the linearly polarized light travelling in such a ``medium'' experiences a rotation of the direction of polarization, which is called \textit{Cosmic Birefringece}. A new way to understand such a phenomenon is as a ``photon chiral memory effect" \cite{Maleknejad:2023nyh}. This memory effect measures the permanent change of the spin angular momentum of electromagnetic fields by chiral symmetry violating processes ($\phi F \Tilde{F}$ in our case) occurring in the bulk. This reveals an interesting connection between the birefringence effect, electromagnetic memory effect, and the asymptotic symmetries/charges at null infinity\footnote{As yet, these effects have been discussed in the context of asymptotically flat space-times. For cosmic birefringence  the curvature of the universe is non-negligible and  should be  taken into account.}~\cite{Maleknejad:2023nyh}.}\\
Within the axion framework, the cosmic birefringence angle depends only on the  field displacement  between the times of emission and  observation  and it is frequency independent \cite{Carroll:1989vb, PhysRevD.43.3789, Harari:1992ea,Lue:1998mq}, which is consistent with  current results \cite{Eskilt:2022cff}. \jo{ The angle of rotation of CMB light} is thus significant when the scalar field has a non-trivial evolution between  LS and today\jo{. In particular, the dominant contribution comes from fields that start oscillating in this time window, corresponding to axion masses of $H_0\sim 10^{-33} \text{eV} \leq m \leq H_{\text{LS}}\sim 10^{-29}$ eV. }  We focus on axion fields produced via the misalignment mechanism with decay constant $f_a$ greater than the energy scale of inflation $H_I$, such that each field is mostly uniform in our observable patch of the Universe \cite{Preskill:1982cy, Abbott:1982af, Dine:1982ah, Marsh:2015xka}. 
\jo{This is the most common scenario discussed in the literature \cite{Pospelov:2008gg,Finelli:2008jv,Panda:2010uq,Lee:2013mqa,Zhao:2014yna,Galaverni:2014gca,Liu:2016dcg,Sigl:2018fba,Fedderke:2019ajk,Fujita:2020aqt,Fujita:2020ecn,Murai:2022zur,Obata:2021nql,Alvey:2021hjp,Gasparotto:2022uqo,Eskilt:2023nxm} to explain the \textit{static and isotropic} nature of the signal, which is generated through the evolution of the axion background. In this case, the unavoidable fluctuations that would be generated during inflation are suppressed by a  factor of $\sim H_I/f_a$ with respect to the isotropic component, in line with the current absence of evidence of an \textit{anisotropic} counterpart \cite{Bortolami:2022whx,Namikawa:2020ffr,SPT:2020cxx}.  Note that in the whole mass range that could give rise to the signal ($10^{-40} \text{eV} \lesssim m \lesssim 10^{-26}$ eV \cite{Fujita:2020aqt}), the axion field can be responsible for dark energy or comprise a subdominant fraction of dark matter. With future experiments, it will be possible to access higher masses by 
examining the different effects on
$\beta$ of the CMB polarization at early and late times \cite{Fedderke:2019ajk}.
The signal could be also generated by axions forming a long-lived network of domain walls or cosmic strings; this scenario would be distinguishable through a characteristic imprint on the \textit{anisotropic} spectrum of cosmic birefringence \cite{Takahashi:2020tqv,Jain:2021shf,Jain:2022jrp,Kitajima:2022jzz,Gonzalez:2022mcx}. }

However, one can argue for the existence of many axion fields or axion-like particles (ALPs), that were dynamical in various stages of cosmic evolution.
In the context of the String Axiverse \cite{Arvanitaki:2009fg, Marsh:2015xka, Mehta:2021pwf}\jo{\cite{Cicoli:2012sz, Svrcek:2006yi, Cicoli_2014}},  many axion fields exist (hundreds or thousands) and are spread over several orders of magnitude in mass\footnote{\jo{Interestingly, cosmic birefringence provides a complementary  view of the Axiverse, compared to other axion-driven phenomena, such as supperadiance, due to the different axion masses involved. However, CMB birefringence requires the axions to couple to the SM photon, which the supperadiance mechanism does not.}}. We thus explore the implications of having $\beta \sim 0.3$ deg in the presence of many dynamical axions or ALPs. 
Previous work has explored the case of a single field \cite{Fujita:2020aqt,Fujita:2020ecn} where constraints on the axion-photon coupling $g_{\phi\gamma}$ are inferred by inverting the relation $\beta=-\frac{1}{2}g_{\phi\gamma}[\phi(t_{\text{LS}})-\phi(t_0)]$, where $t_0$ and $t_{\text{LS}}$ correspond to the present and ``last scattering'' time and the initial condition is fixed by the maximum axion abundance in the relevant mass range \cite{Hlozek:2014lca,Rogers:2023ezo}. In this work, we follow a different approach: we consider the field's initial value and the axion decay constant as random variables drawn from theoretically motivated distributions. We can then calculate the statistical distribution of the resulting birefringence angle as a function of the number of axions and check how many are needed to match the observed signal within one standard deviation, for different distributions of the input parameters (axion mass, decay constant etc).

Interestingly, we find that combining the results for the cosmic birefringence and the axion abundance gives a generic upper bound for the central value of the decay constant, independently of the other parameters for the quadratic potential. Regarding the mass distribution, so far the cosmic birefringence cannot constrain it directly, but we show that the projected abundance at higher masses gives important information on the maximum allowed mass. Moreover, the future measurement of the birefringence from lower multiples, corresponding to photons emitted around reionization \cite{Sherwin:2021vgb, Nakatsuka:2022epj, Galaverni:2023zhv}, can be used as a probe of the mass distribution which is typically uniformly distributed in  logarithmic space.
We also discuss the impact of correlations, especially between $(\phi_{\rm in},f_a)$ and $(m_a,f_a)$, that can emerge from the underlying UV theory \cite{Mehta:2021pwf,Broeckel:2021dpz}. In the second part of the paper, we repeat the study for axions with monodromy potential that have a different monomial dependence for field values larger than some new transition scale and we show how the constraining region changes in these cases.

The paper is structured as follows:
In Section~\ref{sec:model} we introduce the model, describe the single field result for axion-driven CMB birefringence and introduce the notion of a statistical treatment in the presence of many dynamical axion fields in the universe. 
In Section~\ref{sec:cosine} we present the many-field results in the case of the usual cosine potential. In Section~\ref{sec:quadratic} we analyze the case of many axions with independent quadratic potentials. This allows for an in-depth analysis and for several  results to be derived analytically. In Section~\ref{sec:monodromy} we use axion monodromy potentials, focusing on potentials that are asymptotically linear at large field values.
We discuss our findings and propose future directions in Section~\ref{sec:summary}.

\section{Model}
\label{sec:model}

The Lagrangian of an axion-like particle  (ALP) $\phi$ coupled to a $U(1)$ gauge field, in this case the Standard Model photon, can be written as
\beq
\label{eq:lagrangian}
{\cal L} = -{1\over 2} (\partial \phi)^2+V(\phi) -{1\over 4}F^2 -{1\over 4}{g_{\phi\gamma}} \phi F\tilde F \qquad{\rm with}\qquad g_{\phi\gamma}=\frac{\alpha_{\text{EM}}}{2\pi f_a}c \, ,
\eeq
where $c$ is the anomaly coefficient
and $\alpha_{\rm EM}\simeq 1/137$ is the fine-structure constant. This leads to the well-known equation of motion for the two circular polarizations of the gauge field in an expanding universe with a scale factor $a$:
\beq
\ddot A_k^{\pm} + H \dot A_k^{\pm} +\left (
{k^2\over a^2}\mp {k\over a}{g_{\phi\gamma}} \dot\phi
\right )A_k^\pm=0 \, ,
\eeq
where dots denote derivatives  with respect to cosmic time  and $H=\dot{a}/a$ is the Hubble parameter.
It is evident from the dispersion relation $\omega_{\pm}^2\equiv {{k^2\over a^2}\mp {k\over a}{g_{\phi\gamma}} \dot\phi}$ that the two circular polarizations ($\pm$) are affected differently by the rolling pseudo-scalar field $\phi$. This can have interesting phenomenology, including efficient transfer of energy to gauge fields, the fragmentation of $\phi$ and the generation of gravitational waves \cite{Ratzinger:2020oct, Madge:2021abk, Adshead:2015pva, Adshead:2016iae}. 
In the current work, we are instead interested in the different phases of the two polarizations of CMB photons, as they propagate through the Universe under the effect of a slowly rolling ALP \cite{Carroll:1989vb, Harari:1992ea, Pospelov:2008gg}.

The total birefringent angle $\beta$ affecting the CMB photons generated by the intervening axion field is given by the integral over the line of sight of the difference $\omega_{-}-\omega_+$ which is  shown to be \cite{Harari:1992ea} $g_{\phi\gamma}\dot{\phi}/2$ and  can be separated into two parts \cite{Fujita:2020aqt}:       
\begin{align}
&\beta(\hat{n})=\frac{g_{\phi\gamma}}{2}\int^{t_0}_{t_{{\rm LS}}} \dot\phi(\hat{n}) {\rm d}t =\frac{g_{\phi\gamma}}{2}(\Delta{\phi} + \Delta\delta\phi(\hat{n})),\\
& \Delta{\phi} ={\phi}(t_{0})-{\phi}(t_{\text{LS}})\, , \qquad \Delta\delta\phi(\hat{n})= \delta\phi_{0}-\delta\phi_{\text{LS}}(\hat{n}) .
\end{align}
The first contribution comes from the evolution of the background field whereas the second contribution is sourced by the primordial, quantum in origin, perturbations. We consider axion fields with misalignment production and the $U(1)$ symmetry broken before or during inflation, such that during LS the axion fields are homogeneous over the whole sky, except for perturbations of the order $\delta\phi\sim H_I/f_a\ll \phi_{\text{LS}}$ \cite{Kolb:1990vq}. Thus the evolution of the background field and its local perturbations $\Delta\phi$ and $\delta\phi_{0}$ produce a uniform rotation of the polarization plane over the whole sky called \textit{isotropic cosmic birefringence}. On the other hand, the fluctuations at  LS lead to a spatially dependent rotation called \textit{anisotropic cosmic birefringence}, because the corresponding values of fluctuations in various directions of the sky are uncorrelated at distant positions of the LS surface \cite{Caldwell_2011,Greco:2022ufo,Zhai:2020vob,Greco:2022xwj}. In our analysis, we focus on axions with masses $H_0 \leq m \leq H_{\text{LS}}$ because, for \jo{ similar} initial conditions, the field displacement of these fields dominates the signal by orders of magnitude. Indeed, the contribution from lighter fields $m\ll 10^{-33}$ eV is suppressed by $(m/H_0)^2\ll1$\jo{, this is because they are  just starting to roll  and their evolution deviates from the initial value by a factor $(mt_0)^2\ll1$ where $t_0\sim H_0^{-1}$ is the age of the Universe. On the other side,} heavier fields $m\gg 10^{-29}$ eV experience a wash-out effect due to the several oscillations over the finite duration of the LS~\cite{Fedderke:2019ajk}\jo{\cite{Arvanitaki:2009fg}}. In the next section, we briefly review the evolution of a rolling axion field during the matter-dominated era (the complete analysis can be found in Ref.~\cite{Marsh:2015xka}) that corresponds to axions with masses in the window of interest, allowing us to use the corresponding equations in our analysis. Subsequently, we discuss some theoretical motivations for choosing the form of the probability density functions (PDFs) for the axion properties (including the mass and axion decay constant), drawing inspiration from the String Axiverse scenario.

\subsection{ALP background motion  as a source of CMB birefringence}
\medskip

The axion field evolves under the  Klein-Gordon equation in an expanding universe, which in the quadratic approximation of the potential is written as 
\begin{equation}\label{eq:EOMaxionquadratic}
    \ddot{\phi}+ 3H\dot{\phi} + m^2\phi=0 \, .
\end{equation}
If we consider the ansatz $a(t) \propto t^p$, where $p=1/2$ in radiation-domination and $p=2/3$ in matter domination (we assume that the scalar field does not dominate the expansion rate), the scalar field evolution is expressed in terms of Bessel functions as:
\begin{align}\label{eq:Besselsolution}
    &\phi(t)= a(t)^{-3/2}(mt)^{1/2}(AJ_n(mt)+ BY_n(mt))
\end{align}
with $n=(1/2)\sqrt{9p^2-6p+1}$.
We neglect the solution $Y_n$ since it is divergent at early times, whereas we know that the initial value is determined by the misalignment angle. 
The parameter $A$ is fixed by the initial field amplitude. The main features of the evolution are captured by $J_n$, which is approximately constant when $mt\ll1$, meaning that the field is ``frozen'', and it oscillates as $\cos(mt)$ when $mt\gg1$, corresponding to dark matter behaviour. The transition between the two behaviours is well approximated by $mt=1$ or, using $H=p/t$,  $ H_{\text{osc}}=p m$. After the transition, one can easily show that \jo{for $a_{\rm osc}\gtrsim a_{\rm LS}$, i.e. in matter domination,} the field amplitude scales like $\phi(t)\simeq\phi_i \left [{a_{\text{osc}}}/{a(t)}\right ]^{3/2}$ \jo{and $\phi_{\rm osc}\sim \phi_i$}. Thus the axion displacement, in the mass range of interest, is $\Delta\phi\simeq -\phi_{\text{LS}}$ because $\phi_0\ll\phi_{\text{LS}}$ and the latter is related to the present field value as:
    \begin{align}
    &\phi_{\text{LS}}= \left(\frac{a_{osc}}{a_0}\right)^{-3/2}\phi_0\simeq \frac{2}{3}{m\over H_0}\phi_0 \, .
    \end{align}
Because the field is frozen before  LS, we can approximate $\phi_{\text{LS}}\simeq \phi_{\text{in}}$ and relate it to the current abundance as:
\begin{align}\label{eq:Omegaquadest}
        \Omega_\phi  = \frac{\rho_\phi}{\rho_{\text{cr}}}=\frac{1}{3H_0^2M_{{\rm Pl}}^2}\left(\frac{\dot{\phi}^2}{2}+\frac{m^2\phi^2}{2}\right )
        \simeq \frac{m^2\phi^2_0}{6H_0^2M_{{\rm Pl}}^2}\simeq \frac{m^2}{6H_0^2M_{{\rm Pl}}^2}\frac{9H_0^2}{4m^2}\phi^2_{\text{in}}= \frac{3}{8}{\phi^2_{\text{in}}\over M_{{\rm Pl}}^2 }.
    \end{align}
Therefore in this mass regime, the field displacement takes the value
\begin{equation}\label{eq:DeltaH0HLS}
        \Delta\phi\simeq -\phi_{\text{in}} \simeq -\frac{2}{3}\sqrt{6\Omega_\phi}\, M_{\rm Pl}.
    \end{equation}
Note that in this mass regime, the current density is more constrained compared to higher or smaller axion masses from linear cosmology as studied in Ref.~\cite{Hlozek:2014lca} and more recently in Ref.~\cite{Rogers:2023ezo}. In particular, axions \jo{are constrained} to \jo{comprise} about 1\% of the total dark matter abundance, $\Omega_\phi\leq 0.003$, for $ 10^{-33}\text{eV} < m < 10^{-29} \text{eV}$, \jo{which from Eq.~\eqref{eq:Omegaquadest} translates into $\Delta\phi/M_{\rm Pl}\lesssim 0.09$.}  

In the single-field axion case, the birefringence angle is given by:
\begin{equation}\label{eq:birangleesti}
    \beta=g_{\phi\gamma}\frac{\Delta\phi}{2}= \frac{\alpha_{\text{EM}}c}{2\pi f_a} \Big(\frac{\Delta\phi}{2}\Big)\Big(\frac{360}{2\pi}\Big) \, \text{deg}=0.033\Big(\frac{M_{\rm Pl}}{f_a/c}\Big) \Big(\frac{\Delta\phi}{M_{\rm Pl}}\Big)\, \text{deg} .
\end{equation}
Combining Eqs.~\eqref{eq:DeltaH0HLS}  and \eqref{eq:birangleesti}  leads to
\begin{equation}
    \beta\leq0.003\Big(\frac{M_{\rm Pl}}{f_a/c}\Big)\, \text{deg}.
\end{equation}
Using the measurement $\beta_{\rm obs}\sim 0.3$ deg, \jo{ one finds ${f_a/c}\leq 10^{-2}{M_{\rm Pl}}$}, which fits well with the theoretical predictions from string theory \jo{\cite{Cicoli:2012sz, Svrcek:2006yi, Cicoli_2014,Arvanitaki:2009fg}}. We later show how these results extend in the case of a large number of fields.  

\subsection{Statistics of many ALP's}
\label{sec:statisticsALPs}
In this work, we consider the existence of many axions whose parameters are distributed following some probability density function (PDF). These PDFs can be either derived through some  underlying UV theory or taken as phenomenological ``parameters'', which can be constrained through experiments and observations. Indeed, in String Theory, the axions are naturally linked to the geometry of the compactification that determines the distribution of the mass $m$ and decay constant $f_a$ \jo{\cite{Arvanitaki:2009fg, Marsh:2015xka,Acharya:2010zx,Cicoli_2014,Svrcek:2006yi,Demirtas:2018akl}. Generically, instantons generate a potential for the axions that is exponentially suppressed by the instanton action, thus $m\propto e^{-S_{ins}}$. This last quantity is linked to the parameter of compactification which is likely to be uniformly distributed rather than being concentrated around one particular scale \cite{Arvanitaki:2009fg}, giving a uniform mass distribution in logarithmic space (see App E of Ref.~\cite{Marsh:2015xka}).} \jo{This distribution has been found in different compactification scenarios, for example \cite{Acharya:2010zx,Cicoli_2014,Demirtas:2018akl}, and numerically in Ref.~\cite{Mehta:2021pwf}, at least far from the Planck scale. The extremes of the mass distribution are model-dependent, thus we take an agnostic point of view and consider the mass bounded from above by the Planck mass $m_{\rm max}\lesssim M_{\rm Pl}$. Determining minimum mass is more difficult, since it is also connected to  dark energy  \cite{Kaloper:2008qs}. We follow  Refs~\cite{Mehta:2021pwf,Acharya:2010zx,Arvanitaki:2009fg} and choose $m_{\min}\sim H_0$.}

\jo{String theory, in general, predicts a non-flat distribution for the decay constant which has the following parametric dependence $f_a\sim M_{\rm Pl}/S_{ins}$ for a single field \cite{Svrcek:2006yi}.  For the Axiverse, it has been shown that in several cases it follows a Gaussian distribution in logarithmic space with a decreasing mean value for an increasing number of axions \cite{Mehta:2021pwf,Broeckel:2021dpz,Halverson:2019cmy,Mehta:2020kwu}. In particular, in Ref.\cite{Mehta:2020kwu}  it is explained how the log-normal distribution of the decay constant, computed as $f_a^2=m^2/\lambda$, as the product distribution of the mass and the axion self-interaction coupling $\lambda$ that follows a similar distribution along with a strong correlation between the two which make the distribution peak at a certain scale. For comparison, in section \ref{sec:correlatedquadratic} we  discuss the results for the log-normal and log-uniform distribution for the decay constant. }

The initial  value \jo{of an axion} in a cosine potential can be taken to be uniformly distributed in the fundamental domain of the cosine. In a non-periodic potential, like a quadratic or axion monodromy potential, 
de-Sitter fluctuations during inflation can generate a large field value. For simplicity, and in order not to include specific inflationary dynamics in our analysis, 
we assume that the distribution of initial field values follows a Gaussian distribution with a spread 
$\sigma_\phi<M_{\rm Pl}$. 

We explore these distributions in the context of CMB birefringence by varying their mean and standard deviation. Furthermore, we also consider the existence of correlations between the different parameters \jo{, as correlations have been shown to arise in the Axiverse \cite{Mehta:2021pwf,Broeckel:2021dpz}. For example, a correlation between the initial field value and the decay constant can be understood as the initial field value given by the product of the initial misalignment angle and the decay constant $\phi=f_a \theta$. This last quantity differs from the scale entering in the coupling $g_{\phi\gamma}\propto c/f_a$ by the anomaly coefficient $c$, thus we can imagine that in the presence of axion mixing the relation $\phi$ and $f_a$ is not exact, but they can be strongly correlated. Ref.~\cite{Obata:2021nql} provides a concrete example of axion's mixing and the implication for cosmic birefringence.  Similarly, a correlation between the decay constant and the axion mass, which is exact for the QCD axion $m\propto f_a^{-1}$, can arise for string axions when axion's mixing is present \cite{Obata:2021nql,Mehta:2021pwf,Broeckel:2021dpz}.   }

\section{Cosine potential}
\label{sec:cosine}
We start with the cosine potential,
\begin{equation}\label{eq:cospotential}
     V= \sum^N_{i=1} \Lambda^4\left [ 1-\cos\left (\frac{\phi_i}{f_{a,i}}\right )\right ]
\end{equation}
as this is the usual potential for axions, arising from instanton effects (see e.g. Ref.~\cite{Gross:1980br}). Furthermore, it allows us to make contact with the estimate given in Ref.~\cite{Arvanitaki:2009fg}, where $\beta\propto \sqrt{\cal N}$. If there is no mixing between the different axion fields, such that the decay constants appearing in the potential 
$\phi_i/f_{a,i}$ 
and in the couplings $g_{\phi\gamma,i}\sim f_{a,i}^{-1}$ are the same for each axion,  the two cancel in Eq.~\eqref{eq:birangleesti}, since we expect $\Delta\phi_i \propto f_{a,i}$. The
 birefringence angle thus depends only on the initial field position within the fundamental domain of the cosine. Considering $\phi_i/f_{a,i}$ uniformly distributed in the interval $[-\pi,\pi]$, the average value of $\beta$ is zero and its variance is controlled by the variance of the initial field value $\left \langle\left ( \phi_{\text{in},i}/f_{a,i} \right )^2\right \rangle =\pi^2/3$. \jo{ Neglecting  the effects of the anomaly coefficient $c_i\sim\mathcal{O}(1)$ leads to}
\begin{equation}\label{eq:betacos}
\langle \beta^2\rangle \sim
\frac{\alpha_{\text{EM}}^2}{(4\pi)^2}\sum_i^N  \left \langle \left ( \frac{\phi_{\text{in},i}}{ f_{a,i}}\right )^2\right \rangle \sim \left (\sqrt{N}\frac{\alpha_{\text{EM}}}{2\sqrt{3}}\right )^2 \sim \left (0.06\sqrt{N}\right )^2\quad [\text{deg}] \, .
\end{equation}
Requiring that the standard deviation of the birefringence angle matches the observed value $\beta\simeq 0.3$ deg,  leads to $N\simeq 25$ fields, which is a large enough number to justify a statistical treatment. 

\subsection{Anomaly coefficient and correlations}

We go beyond this simple result by introducing 
 the anomaly coefficient $c_i$ that can be different for each axion, drawn itself from a PDF, while we allow it to be correlated to the initial field amplitude $\theta_i\equiv\phi_{{\rm in},i}/f_{a,i}$. The correlation coefficient $\rho$ is defined as 
\beq \rho=\frac{ \langle c_i \theta_i\rangle-\langle c_i\rangle \langle \theta_i\rangle  }{\sigma_c \, \sigma_\theta} \, ,
\eeq 
where $\sigma_c,\sigma_\theta$ are the standard deviations of the two distributions. 
The average value of $\beta$ in the general case is
\beq
\langle \beta \rangle = 0.03 N \langle \theta_i c_i\rangle = 0.03 N \left (
\rho \sigma_c\sigma_\theta + \langle c_i\rangle \langle\theta_i\rangle
\right )
\eeq
We typically consider distributions that are symmetric with respect to zero, meaning that they have zero mean. For example, in order to ``bias'' the distribution of initial field amplitudes $\theta_i$, some parity-violating process must occur in the early universe. This is not impossible, but requires some concrete model-building. If at least one  of the average values of the field amplitude and anomaly coefficient vanishes, then the average value of  $\beta$ scales linearly both with the number of fields, as well as with the correlation coefficient, becoming zero for uncorrelated $c_i$ and $\theta_i$. 
\\
The calculation of the variance of a product of random variables is more complicated; we present all relevant formulas in Appendix~\ref{app:corralation}. In the case of Gaussian random variables with mean $\mu_c$ and $\mu_\theta$ (which we can use as a qualitative guide for our case), the  standard deviation becomes
\beq
\sigma_\beta \equiv \sqrt{ \langle\beta^2\rangle - \langle \beta\rangle^2 } 
= 0.03 \sqrt{N} \sqrt{\mu_c^2\sigma_\theta^2 + \mu_\theta^2\sigma_c^2 + 2\mu_c \mu_\theta \sigma_c \sigma_\theta \rho + \sigma^2_c \sigma^2_\theta(1+\rho^2) } \, .
\eeq
When  both random variables have zero average values, the above equation becomes
\beq
\sigma_\beta =0.03\,  \sqrt{N}  \, \sigma_{\theta}\, \sigma_{c}\, \sqrt{1+\rho^2}
\, .
\eeq
We see that this is similar to Eq.~\eqref{eq:betacos} for $\sigma_\theta, \sigma_c, \rho \sim 1$. 
Most importantly however, the different scaling of $\langle \beta\rangle $ with $N$ and $\sigma_\beta$ with $\sqrt{N}$, means that  $\langle \beta\rangle \gg \sigma_\beta$ in the many-field limit, unless $|\rho|\ll 1$. 
If $\langle c_i\rangle \langle \theta_i\rangle=0$ and $\rho\ne 0$, the mean of the birefringence angle will dominate over its standard deviation. 
This corresponds to the case of aligned initial conditions discussed in  Ref.~\cite{Mehta:2021pwf}, because there is an overall tendency for the axions to give either positive or negative contribution to $\beta$. We do not pursue this case further here and instead focus on independent distributions for the various parameters that determine $\beta$. 
We will return to the subject of correlated random variables in Section~\ref{sec:correlatedquadratic}, for the case of a  quadratic axion potential.

\subsection{\jo{Testing the mass distribution with birefringence tomography}}
\label{sec:cosinetomography}

Subsequently, we can ask about the total number of axions needed to \jo{ explain the observed signal $\beta_{\rm obs}\sim 0.3$ deg for a given PDF of the axion mass.  We consider a  log-flat distribution}
\beq
\begin{split}
& \Hat{f}(m;m_{\rm min}, m_{\rm max}) = {1\over m}{{\text{\jo{$\log_{10}(e)$}}}\over \log_{10} (m_{\rm max}) - \log_{10} (m_{\rm min})} \, ,\quad m_{\rm min}\le m \le m_{\rm max}
\\
& \Hat{f}(m;m_{\rm min}, m_{\rm max}) = 0 \, , \quad\quad\quad\quad\quad\quad\quad\quad\quad\quad\quad\quad{\rm otherwise}
\end{split}
\label{eq:loguniform}
\eeq
\jo{ As discussed in section \ref{sec:statisticsALPs}, we take $m_{\rm max}=M_{\rm Pl}$, thus $\log_{10}(m/\text{eV})\in[ -33\log_{10}(m_{\rm min}/H_0),27]$ leaving some freedom for the lower mass cutoff.
 With this (simple) prior for the mass distribution,
the probability that an axion mass falls within the interesting parameter space for birefringence is given by $\mathcal{P}(-33\leq \log(m/\text{eV})\leq -29)=4/(27+33\log_{10}(m_{\rm min}/H_0))$. Taking $m_{\rm min}\sim H_0$ and $N\simeq 25$ axions in the interesting mass range, as follows from Eq.~\eqref{eq:betacos}, we arrive at to $N_{\rm dec} \simeq 6$ axions per decade of mass and $N_{\text{tot}}\sim 360$ total axions with any mass.} This number is compatible with the typical expectation of a few hundred axions (see e.g. Ref.~\cite{Halverson:2019cmy}).
\jo{However, postulating a large number of axion fields can lead to inconsistencies with the dark matter abundance. Using Eq.~\eqref{eq:Omegaquadest}, which is strictly only true for the quadratic case (and well approximates a cosine, when the field does not start very close to the maximum of the potential), and using $\phi_{\text{in}} ={\cal O}(f_a)$, we estimate that the contribution of each axion  to the dark matter abundance will be $\Omega_\phi ={\cal O}(f_a^2/M_{\rm Pl}^2)$. This is very small for axions with a small decay constant and can be catastrophically large for $f_a={\cal O}(M_{\rm Pl})$.
We will return to this point in Section~\ref{sec:DMabuundancequadratic}, where we compute the ALP contribution to the DM abundance for the case of a quadratic axion potential.} 
Note that one can increase or decrease the inferred number of axions by changing the mean value of the anomaly coefficient, 
which can be computed in a definite axion model. For our purposes, we treat it as a phenomenological parameter, whose properties (value, correlations, PDF) can be constrained from the birefringence measurement. 

In principle, we \jo{can constrain} the mass distribution, since it gives a clear prediction for the ratio of the birefringence angle associated with recombination $\beta_\text{rec}$ and reionization $\beta_\text{rei}$, that can be extracted by studying the angular dependence of the polarization data \cite{Nakatsuka:2022epj,Galaverni:2023zhv}. The angle  $\beta_\text{rec}$ measures how many axions have rolled to the minimum of their respective potential between recombination and the present time whereas  $\beta_\text{rei}$ is only sensitive to those axions that started rolling after reionization ($H_{\text{rei}}\sim 10^{-31}$ eV). Therefore the ratio between the two gives information about the relative amplitude of the mass parameter space explored in the two time intervals:
\begin{equation}
    \frac{\beta_\text{rei}}{\beta_\text{rec}}\simeq\sqrt{\frac{N_{\text{tot}}\mathcal{P}(-33\leq \log_{10}(m/\text{eV})\leq -31)}{N_{\text{tot}}\mathcal{P}(-33\leq \log_{10}(m/\text{eV})\leq -29)}}=\sqrt{\frac{2}{4}}\simeq 0.7,
    \label{eq:cosine_beta_ratio}
\end{equation}
whereas in the ``aligned" case \jo{${\beta_\text{rei}} \simeq 0.5 {\beta_\text{rec}}$.  Note that although the result depends on there being a large enough number of axions in the relevant mass window to justify the statistical treatment, it does not depend on the lower-mass cutoff of the mass prior and therefore is formally independent of the total number of axions across all masses. 
}

We can probe the mass distribution of axions using Eq.~\eqref{eq:cosine_beta_ratio} by assuming that the mass distribution over the interesting range does not follow Eq.~\eqref{eq:loguniform}, but instead has a preference for larger or smaller masses. A phenomenological distribution with this property is 
\beq
\begin{split}
& \Hat{f}(m;m_{\rm min}, m_{\rm max}) \sim {1\over m}{{\text{\jo{$\log_{10}(e)$}}}[\log_{10}(m)]^{n_m} \over \log_{10}(m_{\rm max}) - \log_{10}(m_{\rm min})} \, , m_{\rm min}\le m \le m_{\rm max}
\\
& \Hat{f}(m;m_{\rm min}, m_{\rm max}) = 0 \, , \quad\quad\quad\quad\quad\quad\quad\quad\quad\quad\quad{\rm otherwise}
\end{split}
\label{eq:loguniform_tilt}
\eeq
where $n_m$ is a dimensionless number. For \jo{$n_m=0$} we recover the log-uniform distribution. Note that in Eq.~\eqref{eq:loguniform_tilt} we neglect the overall normalization for simplicity and because it drops out of Eq.~\eqref{eq:cosine_beta_ratio}. Using this ``tilted log-uniform'' distribution, we can compute $ {\beta_\text{rei}}/{\beta_\text{rec}}\simeq 2^{-(n_m+1)/2}$. For example a $\pm 20\%$ difference from the result of the log-uniform distribution requires a tilt of $n_m\simeq -0.5$ or $n_m\simeq 0.7$. 
A larger difference from log-uniform will arise if the mass distribution has a stronger scale dependence, e.g. if it follows a log-normal distribution centered around some preferred mass range. 
Of course,  this result can be degenerate with changes in the potential or introduction of correlations between the various parameters of the model, such as the mass, decay constant and anomaly coefficient (see e.g. Section.~\ref{sec:quadratictomography}). \jo{However Eq.~\eqref{eq:cosine_beta_ratio} provides a clean expectation for the log-flat prior of the axion mass. A future detection that is inconsistent with this result would  rule out this simple Axiverse model. }

\section{Quadratic potential}
\label{sec:quadratic}

We now move to the case of axions with a quadratic potential. This simplification,  allows us to derive several results analytically and
gain physical intuition by testing a variety of PDFs.

\subsection{Fixed decay constant $f_a$}

In the case of a quadratic potential, the total birefringence is given by the sum of the individual contributions \jo{of axions in the mass range $-33\leq \log_{10}(m/\text{eV})\leq -29$}, as given by Eq.~\eqref{eq:birangleesti}:
\begin{equation}
    \beta = \sum_{i=1}^N 0.33  \Big(\frac{\Delta\phi_i}{M_{{\rm Pl}}}\Big)\Big(\frac{M_{{\rm Pl}}/f_a}{10}\Big)\,  [\text{deg}] \,  ,
\end{equation}
\jo{where we reabsorbed the anomaly coefficient into $f_a$.}
By fixing the axion decay constant and taking the field amplitude as a random variable with a PDF that is symmetric around zero, the standard deviation of the birefringence angle is 
\begin{equation}\label{eq:birN}
\sigma_\beta\equiv \sqrt{\langle \beta^2 \rangle} =0.33  \sqrt{N}  \frac{\sigma_\phi}{M_{{\rm Pl}}}\left (\frac{M_{{\rm Pl}}/f_a}{10}\right ) \,  [\text{deg}] \, . 
\end{equation}
Therefore, by using the observed value $\beta_{\rm obs} \simeq 0.3$ deg and equating that to the standard deviation $\sigma_\beta \simeq \beta_{\rm obs} $, we find a simple dependence of the number of axions on the standard deviation of the initial field value $\sigma_\phi$ and the fixed decay constant $f_a$:
\begin{equation}\label{eq:N}
    N=100\left(\frac{f_a}{ \sigma_\phi}\right )^2. 
\end{equation}

From this expression, we see that the number of axions depends quadratically on the decay constant and the standard deviation of the initial field value, thus by varying these parameters the required number of axions changes significantly. Finally, in order for our statistical analysis to be viable, we must require that $N\gg 1$ which gives $f_a\gg \sigma_\phi/10$.

It is worth remembering that the (standard deviation of the) birefringence angle 
scales as $\sqrt{N}$, where $N$ is the number of axions\footnote{This is not true for the aligned axion case, which we treat separately.} (with $N\gg1$), whereas for the  axion dark matter abundance \jo{the corresponding scaling is} $\Omega_\phi \propto N$. 
This leads to a constraint on the number of axions predicted from the measurement of $\beta$, so as not to exceed the allowed DM abundance. Even in the simplest case with fixed $f_a$, we derive non-trivial constraints. Extending Eq.~\eqref{eq:Omegaquadest} for the case of $N$ axions, the total axion abundance is:
\begin{equation}\label{eq:NAbundance}
    \Omega_\phi=\frac{3}{8}\sum_{i=1}^N\frac{ \phi_{\text{in},i} ^2}{M_{{\rm Pl}}^2}=\frac{3}{8}\frac{\sigma^2_\phi}{M_{{\rm Pl}}^2}\left(\sum_{i=1}^N\frac{\phi^2_{\text{in},i} }{\sigma^2_\phi}\right) .
\end{equation}
The sum of the squares of $N$ independent Gaussian variables with zero mean is a $\chi_N^2-$distribution with $N$ degrees of freedom, where both the mean $\langle \chi^2_{N}\rangle=N $ and the variance (which is equal to $2N$)  scale with the number of degrees of freedom.\footnote{Note that we later use the half-normal distribution for the field modulus with mean $\mu\ne0$,  thus the mean of the $\chi^2$ shifts as $\langle \chi^2_{N}\rangle=N+\lambda$ where $\lambda=\sum_i^N\mu^2_i$ is the non-centrality parameter, which in our case will be only a small correction.}
The mean value of the total abundance becomes: 
\begin{equation}\label{eq:axionabundance}
    \langle\Omega_\phi\rangle\simeq\frac{3}{8}\frac{\sigma^2_\phi}{M_{{\rm Pl}}^2}N_. 
\end{equation}
The combination $\sigma^2_\phi N$ can be substituted using Eq.~\eqref{eq:N}, leading to:
\begin{equation}\label{eq:omegavsfa}
    \Omega_\phi\simeq\frac{3}{8}\Big(\frac{10f_a}{M_{{\rm Pl}} }\Big)^2\simeq37.5\Big(\frac{f_a}{M_{{\rm Pl}} }\Big)^2.
\end{equation}
Requiring $\Omega_\phi\leq 0.003$ (1\% of the DM abundance in this mass range \cite{Rogers:2023ezo}), we find a constraint for the decay constant $f_a\leq2.2\times10^{16}$ GeV. It is worth comparing this to the constraint on the coupling $g_{\phi\gamma}\gtrsim10^{-20}$ GeV derived in Ref.~\cite{Fujita:2020ecn} using a single rolling axion in a quadratic potential, which gives $f_a\lesssim 10^{17} \, {\rm GeV}$. We see that the many-axion analysis leads to a \jo{somewhat} stricter constraint on the axion decay constant. 

Let us now briefly discuss the ``aligned case'', where the distribution of initial field values has a non-zero mean. Assuming that the mean dominates over the standard deviation, the birefringence becomes
\beq
\langle \beta \rangle = 0.33 N \frac{M_{\rm Pl}/f_a}{10}
\frac
{\langle \phi_{\text{in}} \rangle}{M_{\rm Pl}} \,  [\text{deg}] \, ,
\eeq
leading to $N = 10f_a / \langle \phi_{\text{in}} \rangle$, by
requiring that $\beta_{\rm obs}\sim0.3$ deg.
In order for the standard deviation of the birefringence angle to be negligible compared to its mean, we require
$\sigma_\phi \ll \sqrt{N} \langle \phi_{\text{in}} \rangle = \sqrt{10 f \langle \phi_{\text{in}} \rangle}$. 
Similarly, the average value of the axion abundance is
\beq
\langle \Omega_\phi \rangle = \frac{3}{8} \sum_{i=1}^N\frac{\phi^2_{\text{in},i} }{M_{\rm Pl}^2}
= \frac{3}{8} N 
\frac{\langle \phi^2_{\text{in}} \rangle}{M_{\rm Pl}^2}
= \frac{3}{8} N 
\left (\frac{\sigma_\phi^2 }{M_{\rm Pl}^2}
+
\frac{\langle \phi_{\text{in}} \rangle^2}{M_{\rm Pl}^2}\right)
\eeq
By using the relation for the birefringence, we get
\beq
\langle \Omega_\phi \rangle
=\frac{3}{8} N 
\frac{\sigma_\phi^2 }{M_{\rm Pl}^2}
+
\frac{300}{8N}\frac{f_a^2}{M_{\rm Pl}^2} 
\eeq
We can derive a loose bound, by only considering the second term, leading to $\sqrt{N} > 100 f_a/M_{\rm Pl}$ or equivalently
$
f_a \langle \phi_{\text{in}} \rangle  <10^{-3}M_{\rm Pl}^2
$. The last relation is a condition on the geometric mean of the axion decay constant and the alignment, as given by the average field value.
Another interesting limit is one where all dynamics is controlled by a single scale, the axion decay constant. By taking $\langle \phi_{\text{in}} \rangle \sim f_a\sim \sigma_\beta$, the mean birefringence is larger than its standard deviation, by a factor of $3$, so our assumption for neglecting the standard deviation is borderline valid. The constraint in this case, arising from avoiding overproduction of axion dark mater, is
$f\lesssim 0.01 M_{\rm Pl}$.

The main result of this section can be summarized as follows: \jo{When $\phi_{{\rm in},i}$ has no preferred sign}  the DM abundance gives a maximum effective displacement \jo{of the axion field $\Delta\phi^2\simeq \sum_i\phi^2_{\text{in},i}$} that, \jo{combined with Eq.~\eqref{eq:birN},  leads to}  an upper bound on the decay constant given by Eq.~\eqref{eq:omegavsfa}\footnote{ In the case of dark energy the maximum displacement was set from the upper bound on the equation of state, whereas the abundance was used to fix the initial field value \cite{Gasparotto:2022uqo}.}.
This relation will change if we change the relation between the field value and the abundance, as we see in Section~\ref{sec:monodromy} for the family of axion monodromy potentials.

\subsection{Projected abundance to higher masses}
\label{sec:DMabuundancequadratic}
\begin{figure}
    \centering
    \includegraphics{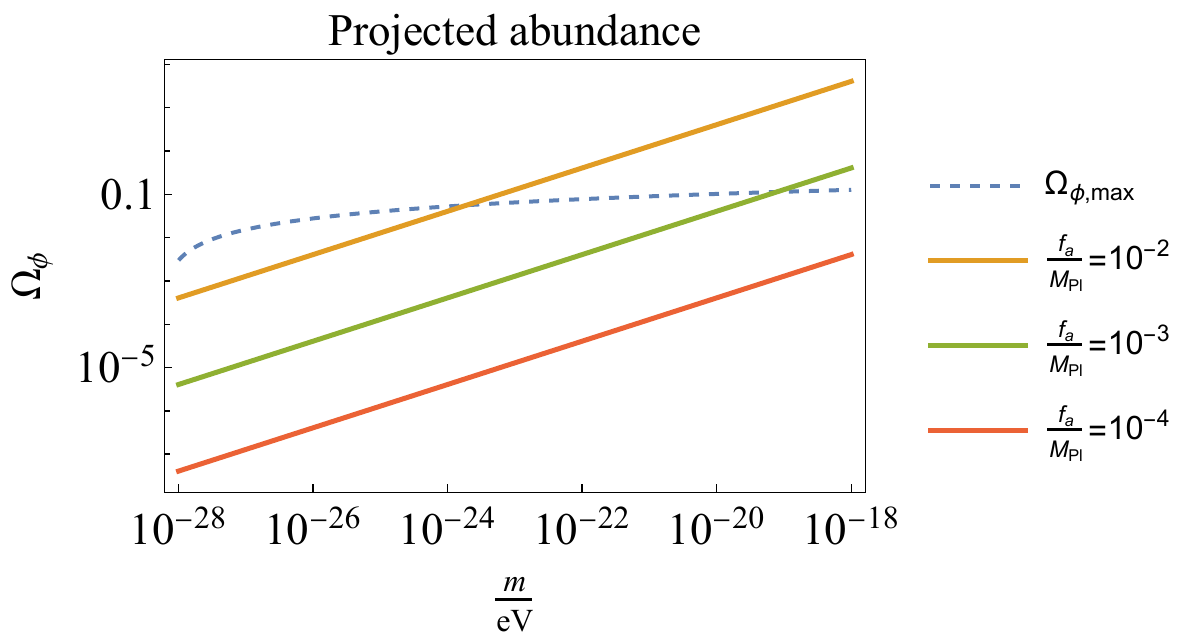}
    \caption{\small The projected abundance for each decade in mass given by Eq.~\eqref{eq:ABmassdecade}. The maximum masses corresponding to the values $f_a/M_{\rm Pl}=\{10^{-2},10^{-3},10^{-4}\}$ shown in the figure are $m_{\text{max}}=\{10^{-24},7\times10^{-20},10^{-15}\}$ eV.}
    \label{fig:ProjectedAB}
\end{figure}
In the axiverse picture, it's interesting to ask whether the expected number of axions we infer from the birefringence measurement, given in Eq.~\eqref{eq:birN}, is consistent with the abundance constraints at higher masses. For instance, we can test the assumption that the same dynamics, which sets the distribution of the initial field displacement and decay constant, also determines the axion abundance at higher masses. For axions with masses $m>H_{\text{eq}}\sim 10^{-28}$ eV, the present abundance for a single axion field is given by \cite{Hlozek:2014lca}
\begin{equation}\label{eq:abhigherm}
    \Omega_\phi=\frac{1}{6}(9\Omega_r)^{3/4}\Big(\frac{m}{H_0}\Big)^{1/2}\Big(\frac{\phi_\text{in}}{M_{\rm Pl}}\Big)^2 ,
\end{equation}
where $\Omega_r$ is the present radiation density. Eq. \eqref{eq:abhigherm} leads to the following abundance for each decade in mass
\begin{equation}\label{eq:ABmassdecade}
    \langle\Omega_\phi(m)\rangle=\frac{1}{6}(9\Omega_r)^{3/4}\Big(\frac{m}{H_0}\Big)^{1/2}N_{\text{dec}}\Big(\frac{\sigma_\phi}{M_{\rm Pl}}\Big)^2 
\end{equation}
which, using $N_{\rm dec}=25 (f_a/\sigma_\phi)^2$,  depends solely on the axion decay constant. We can check that the average abundance, given in Eq.~\eqref{eq:ABmassdecade}, does not exceed the current cosmological upper bound on the axion abundance reported for example in Refs.~\cite{Rogers:2023ezo} and \cite{Kobayashi:2017jcf}. In Figure.~\ref{fig:ProjectedAB} we show the projected abundance and the corresponding upper limit computed as the linear interpolation, in decade of mass, between the following values:  
\begin{equation}
    \frac{\Omega_\phi}{\Omega_{\text{DM}}}\Big|_{\rm max}=\begin{cases}
        0.01 & m\simeq 10^{-28} \text{eV}\\
        0.3 & m\simeq 10^{-21} \text{eV} \\
    \end{cases} .
\end{equation}
The first (stronger) constraint, for axions with $m\leq 10^{-28}$ eV, comes from CMB and Large Scale Structure analyses, which relaxes for $m\gtrsim 10^{-25}$ eV \cite{Rogers:2023ezo}. At higher masses,  the constraint from the Lyman-$\alpha$ forest becomes stronger, which bounds the axion to be at most 30\% of the total dark matter abundance for $m<10^{-21}$ eV \cite{Kobayashi:2017jcf}.

It is interesting to note that the assumption of a unique decay constant for each decade in mass is in conflict with the expectations shown in Figure~\ref{fig:ProjectedAB}, where the corresponding abundance exceeds the cosmological bounds. For instance, as shown in Figure~\ref{fig:ProjectedAB}, this occurs at $m\sim 10^{-24}$ eV for $f_a \sim 10^{16}$ GeV, whereas the abundance is consistent up to $m\sim 10^{-20}$ \jo{eV} for $f_a \lesssim 10^{15}$ GeV. This would indicate, for example, that axions in different mass ranges have different decay constants or that they have different production mechanisms. 
These upper bounds assume that we can do statistics within each decade of mass, which is a rather fine-tuned version of the axiverse. A better estimator is the sum of all axions, whose masses are drawn from a distribution
\beq
\Omega_{\phi,{\rm tot}} = \sum_i \Omega_{\phi,{i}}
= \sum _i \frac{1}{6}(9\Omega_r)^{3/4}\Big(\frac{m_i}{H_0}\Big)^{1/2}\Big(\frac{\phi_{\text{in},i}}{M_{Pl}}\Big)^2.
\eeq
We take the average value of this, leading to
\beq
\langle \Omega_{\phi,{\rm tot}} \rangle = N 
\frac{1}{6}(9\Omega_r)^{3/4}\frac{1}{\sqrt{H_0}} \frac{1}{ M_{\rm Pl}^2} \left \langle \sqrt{m_i}\right \rangle \left  \langle {\phi_{\text{in},i}}^2 \right \rangle 
\eeq
where we took the mass and initial field amplitude to be uncorrelated. Taking $\langle \phi_{\text{in},i}\rangle=0$ implies that $\langle \phi_{\text{in},i}^2\rangle=\sigma_\phi^2$. Considering again a log-uniform distribution for the masses, it is straightforward to compute 
\beq
\left \langle \sqrt{m_i} \right \rangle  = \int_{m_{\rm min}}^{m_{\rm max}} \frac{\sqrt{m}}{m}
\frac{dm}{\log m_{\rm max}/ m_{\rm min}} \simeq \frac{2\sqrt{m_{\rm max}}}{\log m_{\rm max}/ m_{\rm min}}
\, ,
\eeq
where we took $m_{\rm max}\gg m_{\rm min}$.
If we assume that the mass distribution is log-uniform up to some large cut-off, we can compute the upper value of this cutoff $m_{\rm max}$ from the constraint on $\Omega_\phi$. The total number of axions is also related to this cutoff, since $N_{\rm tot} = N_{\rm dec} \log_{10}m_{\rm max}/ m_{\rm min}$. Interestingly, this makes the final expression for the dark matter abundance depend on $m_{\rm max}$ only through the square root, as
\beq
\label{eq:abhighermTOT}
\langle \Omega_{\phi,{\rm tot}} \rangle = N_{\rm dec} 
\frac{2}{\log 10}\frac{1}{6}(9\Omega_r)^{3/4}\sqrt{\frac{ {m_{\rm max}}}{{H_0}}} \frac{\sigma_\phi^2}{ M_{\rm Pl}^2} 
=
 25 
\frac{2}{\log 10}\frac{1}{6}(9\Omega_r)^{3/4}\sqrt{\frac{ {m_{\rm max}}}{{H_0}}} \frac{f_a^2}{ M_{\rm Pl}^2} .
\eeq
Therefore, the final axion abundance is determined by the product of $(f_a/M_{\rm Pl})^2$ and $\sqrt{m_{\rm max}/H_0}$. Note that Eq.~\eqref{eq:abhighermTOT} is almost identical to Eq.~\eqref{eq:ABmassdecade} with the substitution of $m_{\rm max}$, indicating that the total abundance is dominated by the more massive axions, as shown in Figure \ref{fig:ProjectedAB}. Indeed, because of the $\Omega_\phi\sim\sqrt{m}$ dependence, the contribution from lighter axions becomes quickly negligible for a fixed decay constant.

\subsection{Joint PDF  for the decay constant and the initial displacement}
\label{sec:correlatedquadratic}

We now add one more source of complexity in our calculation, by introducing a distribution for the decay constant $\Hat{f}({f_a})$ and a possible correlation \jo{ with} the initial field amplitude $\phi_{\text{in}}$.
\jo{Keeping the  axion  mass in the range $-33\leq \log_{10}(m/\text{eV})\leq -29$ and} expressing 
\jo{$\phi_{\rm in}$} in units of $M_{\rm Pl}$, the total birefringence angle is given by
\begin{equation}\label{eq:betabidis}
    \beta=-\sum_{i=1}^N \text{sgn}(z_i) 0.033\times 2.453\times 10^{18}\left(\frac{\phi_{\text{in},i}}{M_{\rm Pl}}\right)\left(\frac{\text{GeV}}{f_{a,i}}\right) \quad  [\text{deg}],
\end{equation}
where $z$ is a random variable uniformly distributed between $[-1,1]$. 

\begin{figure}
    \centering
    \begin{minipage}[b]{0.45\linewidth}
    \includegraphics[width=7.5cm]{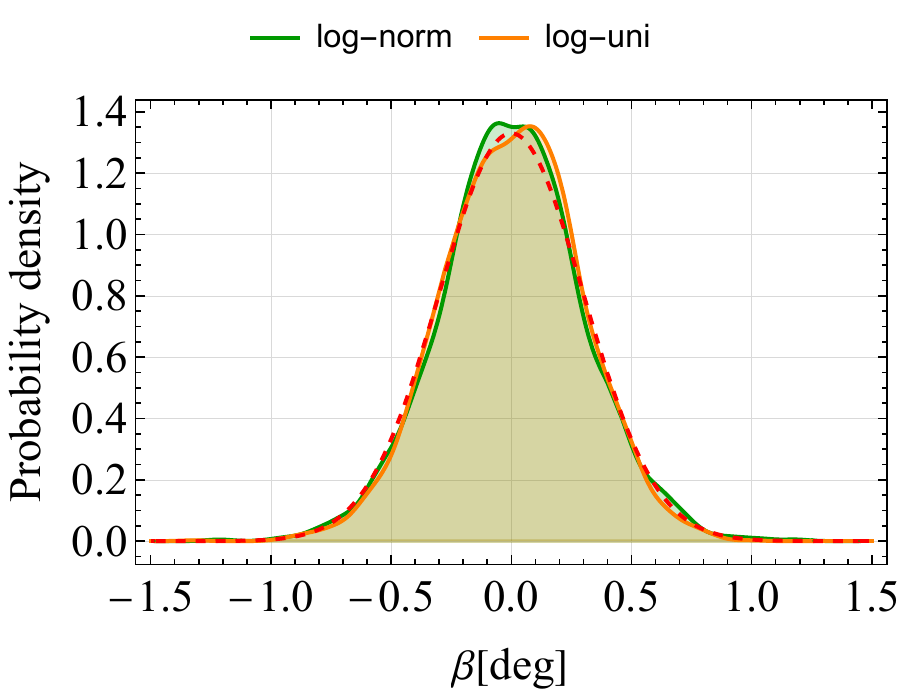}
    \end{minipage}
    \quad
    \begin{minipage}[b]{0.45\linewidth}
    \includegraphics[width=7.5cm]{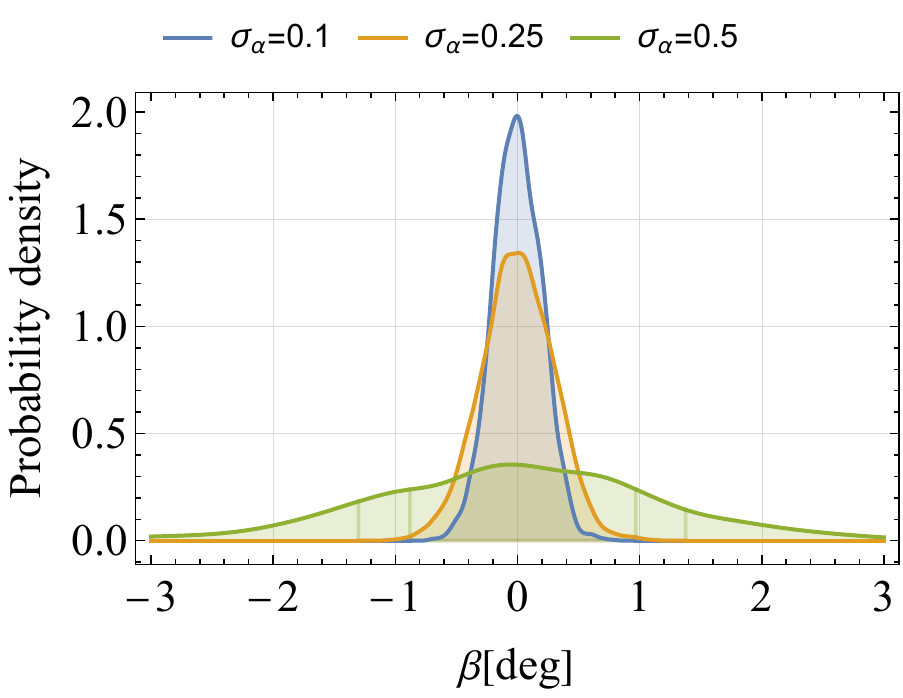}
    \end{minipage}
    \caption{\small \textit{Left}: The
  Probability Density Function of the 
    birefringence angle for the log-normal and log-uniform \jo{marginal} distribution of the decay constant. In both cases we take $N=50$, $\sigma_\phi\simeq 10^{-2}$, $\mu_a=16.5$ and $\sigma_a=0.25$ and the resulting standard deviation of the birefringence angle is $\sigma_\beta\sim 0.3$ deg. \jo{In red we show the Gaussian distribution with $\sigma=0.3$ that  captures the emergent birefringence distribution.}
    \textit{Right}: The birefringence PDF for the log-normal distribution with different values of $\sigma_a$. \jo{We present the curves as smooth instead of histograms to make them visually easier to compare.} }
    \label{fig:birrpdf}
\end{figure}

We want to focus on the size of the initial field value rather than the sign, especially when we consider correlations, thus we take \jo{as marginal distribution for $\phi_{\text{in}}$ } a half-normal distribution \jo{which is }defined only for positive values, whereas \jo{for} $f_a$  \jo{we choose} a log-normal distribution:  
\begin{align}\label{eq:PDFs}
&\Hat{f}(\phi_{\text{in}}) ={1\over \mu_\phi} e^{-{\phi_{\text{in}}^2 \over \pi \mu_\phi^2}}\  \quad{\rm with} \qquad \phi_{\text{in}}\geq0 \\
&\Hat{f}(f_a)=\frac{1}{f_a\sigma_a\sqrt{2\pi}}e^{-\frac{(\ln{f_a}-\mu_{a})^2}{2\sigma_a^2}} \, 
\end{align}
with corresponding average values $\langle  \phi_{\text{in}}\rangle =\mu_\phi$ and $\langle f_a\rangle=\exp{(\mu_{a}+\sigma_a^2/2)}$\footnote{Since we want to show the results in terms of  \jo{ $\log_{10}\langle f_a\rangle$}, for computational convenience, we shift the input parameter as $\mu_a\rightarrow\mu_a-(\sigma_a/\log_{10}(e))^2/2$ \jo{where  $\log_{10}(e)$ is due to our choice of the base of 10 in $\mu_a$}.}. Despite focusing on these \jo{marginalised} distributions, we checked that the resulting distribution of the total birefringence angle does not change significantly if we take the distribution of the decay constant uniform in logarithmic space\footnote{When we adjust the extremes of the uniform distribution as $\log_{10}(f_{a,\text{max/min}}/\text{Gev})=\mu_a\pm\sqrt{12}/2\sigma_a/\log_{10}(e)$.}, as shown in Figure \ref{fig:birrpdf}. Of course, this result depends on the dispersion of $f_a$ and, for increasing values of $\sigma_a$, the result starts diverging for the normal and uniform distribution as can be seen in Figure~\ref{fig:comparison Sigma}, because the ``tails'' of the normal become flatter  and the signal is dominated by the low values of $f_a$.  
In what follows, we take $\sigma_a=0.25$, as the results are reasonably similar for the two distributions, and give an $O(1)$  dispersion for $ \log_{10}(\langle f_a/{\rm GeV}\rangle)$.
Appendix~\ref{app:copulas} describes the construction of joint probability density functions and
some examples of PDFs are shown in Figure \ref{fig:PDFVisual}. 

\begin{figure}
    \centering
    \begin{minipage}[b]{0.5\linewidth}
    \includegraphics[width=8cm]{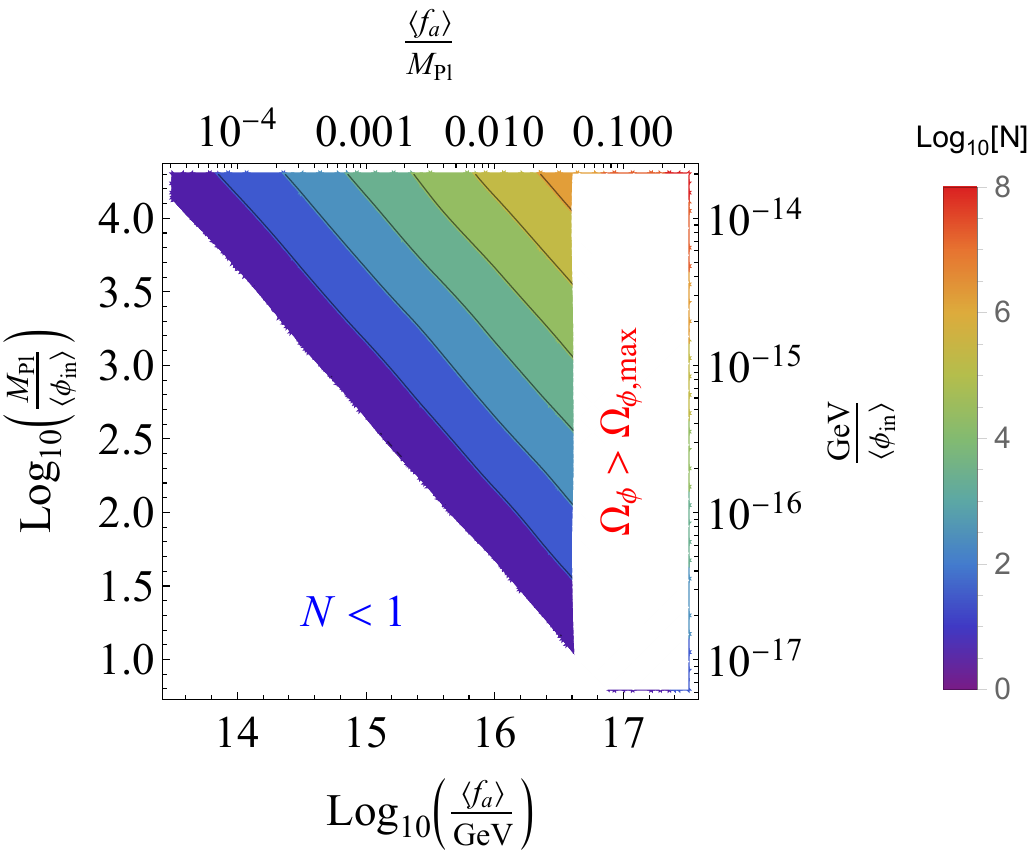}
    \end{minipage}
    \quad
    \begin{minipage}[b]{0.45\linewidth}
    \includegraphics[width=8cm]{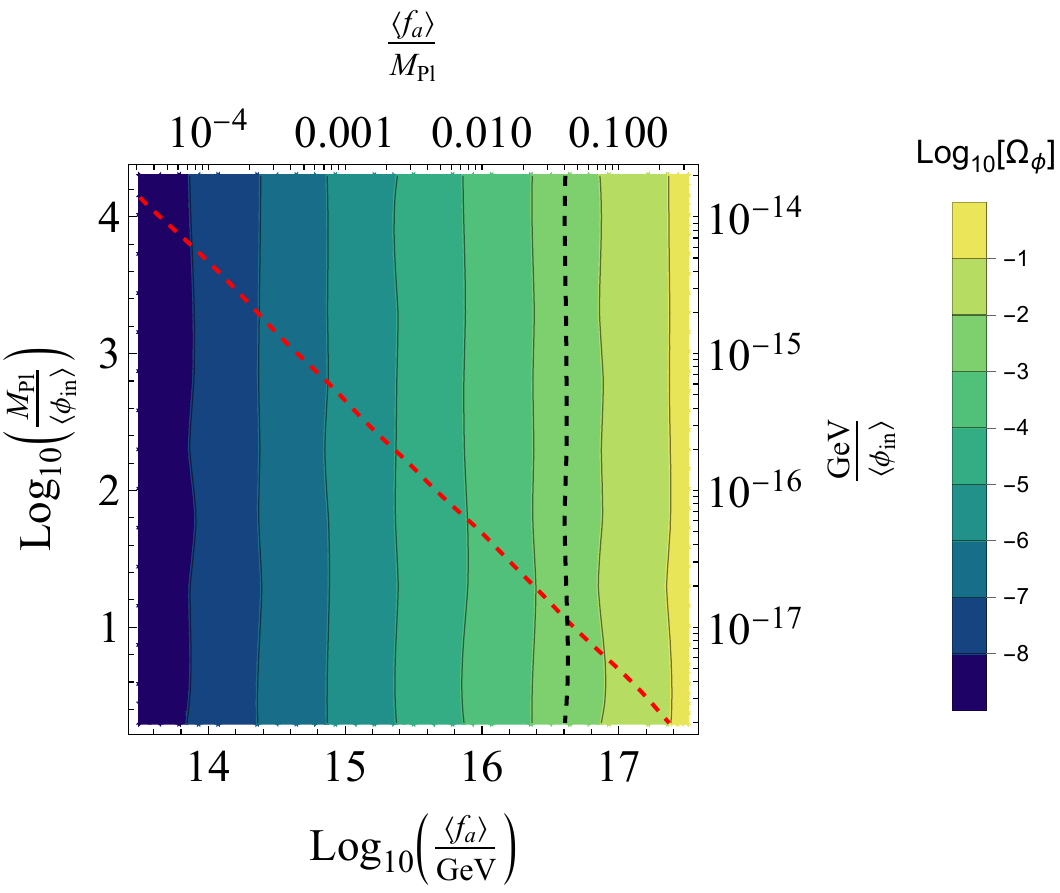}
    \end{minipage}
    \caption{\small \textit{Left:} Number of axions needed to saturate $\sigma_\beta\sim 0.3$ deg as a function of the $\mu_\phi$ and $\mu_a$ parameters of the PDFs \eqref{eq:PDFs} \jo{in the case of zero correlation $\rho=0$}. Blank regions are excluded because $N<1$ whereas the high $f_a$ region is excluded from abundance constraints. \textit{Right:} Abundance of the total axions whose number comes from the left panel. \jo{The red dashed line corresponds to $N=1$ coming from the left panel whereas the black line corresponds to $\Omega_{\phi,{\rm max}}=3\times 10^{-3}$.}  
    Note that the ``waviness'' of the lines in both panels can be attributed to sampling noise.
    }
    \label{fig:zerocorr}
\end{figure}

\begin{figure}
    \centering
    \begin{minipage}[b]{0.5\linewidth}
    \includegraphics[width=8cm]{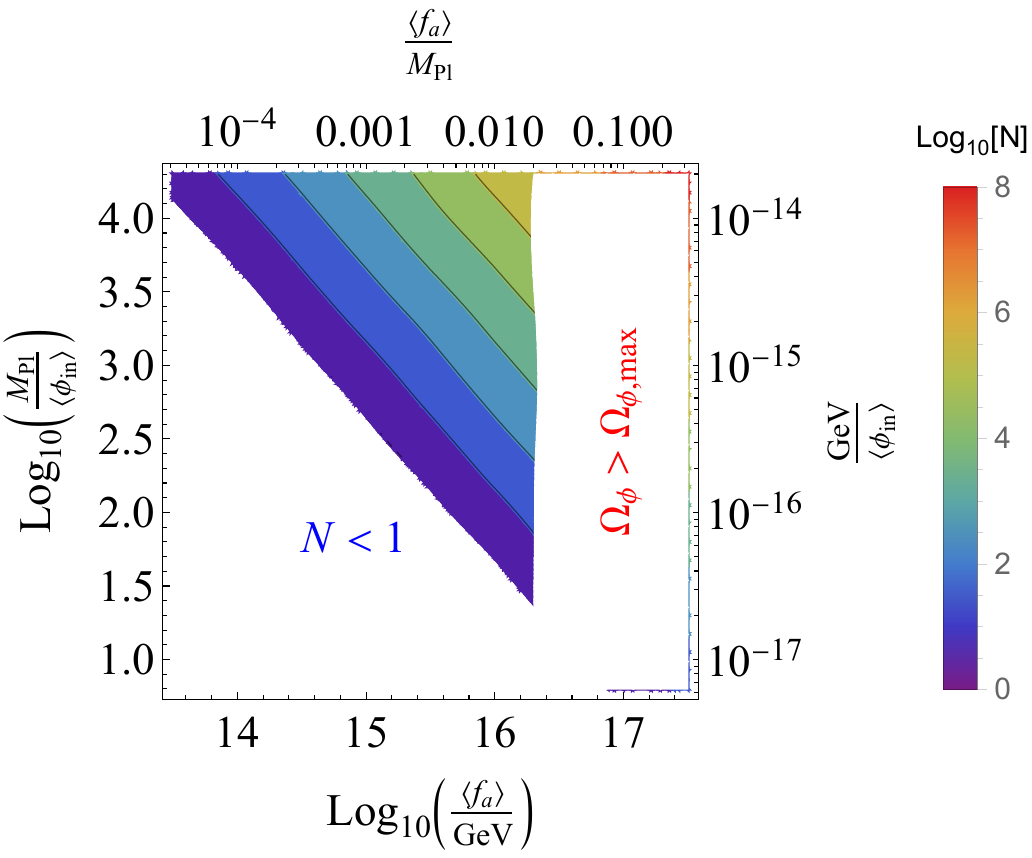}
    \end{minipage}
    \quad
    \begin{minipage}[b]{0.45\linewidth}
    \includegraphics[width=8cm]{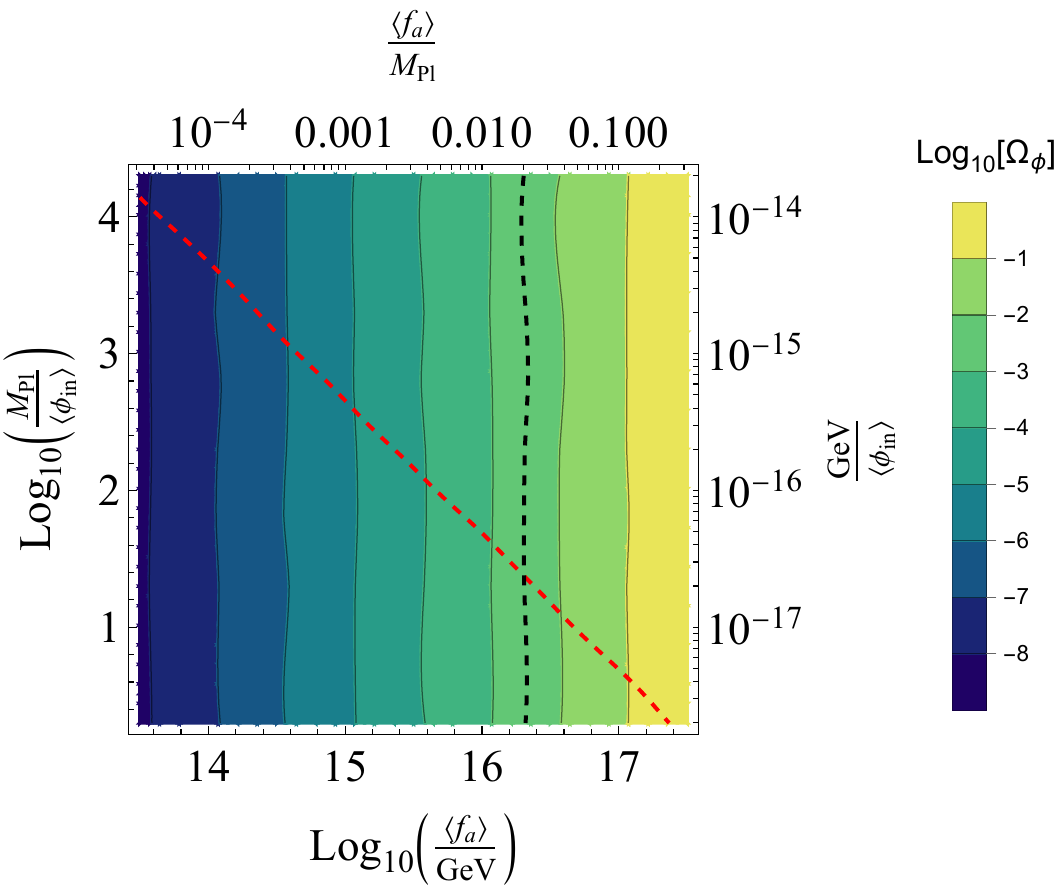}
    \end{minipage}
    \caption{\small Same as Figure~\ref{fig:zerocorr}, but for positive correlation $\rho=0.9$ between  $\phi_{\rm in}$ and $f_a$.
    }
    \label{fig:poscorr}
\end{figure}

\begin{figure}
    \centering
    \begin{minipage}[b]{0.5\linewidth}
    \includegraphics[width=8cm]{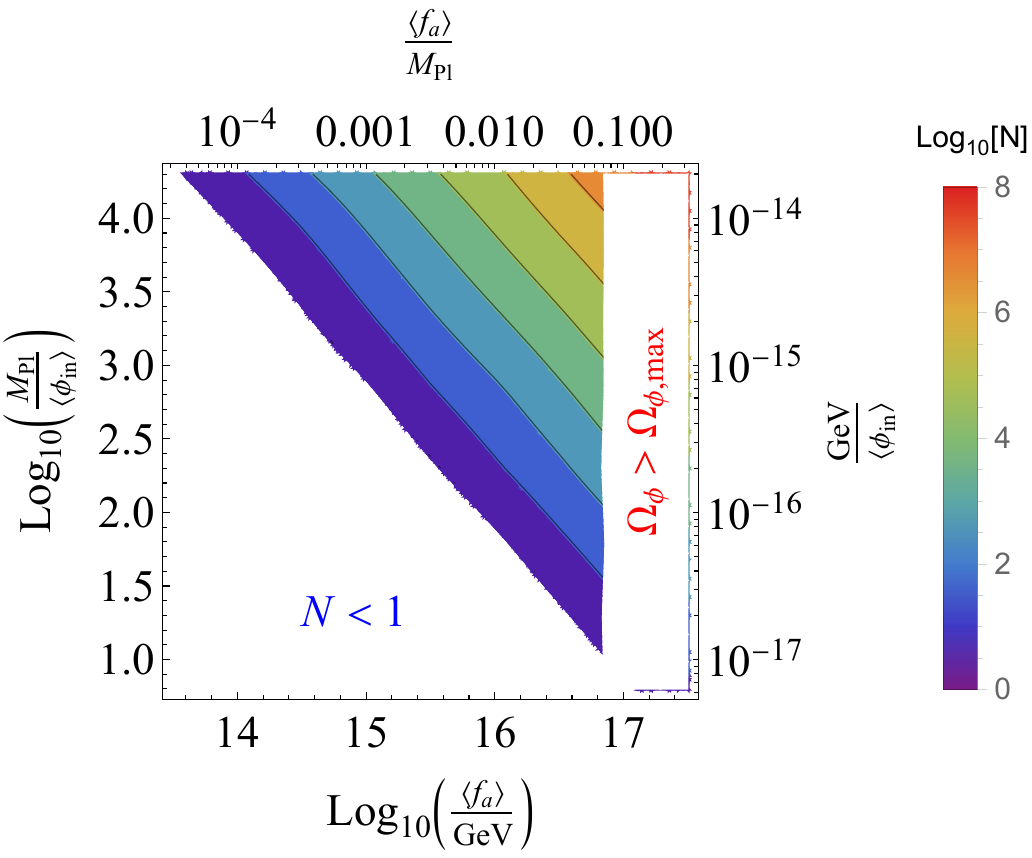}
    \end{minipage}
    \quad
    \begin{minipage}[b]{0.45\linewidth}
    \includegraphics[width=8cm]{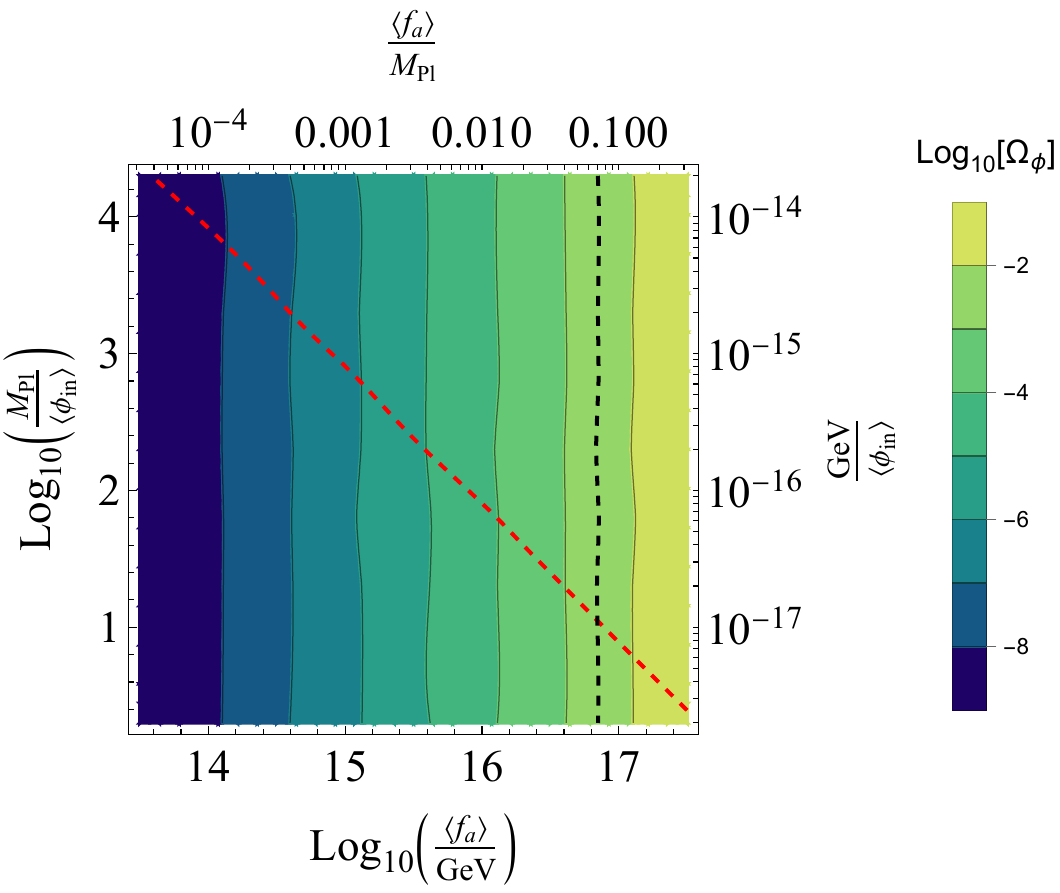}
    \end{minipage}
    \caption{\small Same as Figure~\ref{fig:zerocorr}, but for negative correlation $\rho=-0.9$ between  $\phi_{\rm in}$ and $f_a$.}
    \label{fig:negcorr}
\end{figure}

The first important result is shown in Figure~\ref{fig:zerocorr}, which displays the number of axions needed to match $\beta_{\rm obs}\sim0.3$ deg as a function of the mean values of the distributions of $(f_a,\phi_{\rm in})$ and the corresponding total abundance. We explain the details of the numerical simulations in Appendix~\ref{app:numdet}. The main message is that the number of axions is very sensitive to the mean values of $(\phi_{\rm in},f_a)$, exhibiting an approximately quadratic dependence as follows from Eq.~\eqref{eq:N}, and equal-$N$ lines positively correlate the two parameters. The bottom-left corner of the parameter space, corresponding to the low decay constant and large initial field value, leads to $N<1$, meaning that \jo{ not even one axion is allowed in this regime. Note that for the statistical treatment to be valid one should demand $N\gtrsim 10$, which corresponds to the second contour line in Figure~\ref{fig:zerocorr} and in the following ones.}  Interestingly, we find that the constraints on the abundance, whose dependence on the number of axions is shown on the right panel of Figure~\ref{fig:zerocorr}, translate into a constraint on the decay constant which is independent of the initial field value, as was previously found for the fixed $f_a$ case. Figures~\ref{fig:poscorr} and \ref{fig:negcorr} show the same results when positive and negative correlations are introduced between the two input parameters.  

Even in these cases, the results do not qualitatively change and
the equal abundance contours remain straight, but their position shifts to the left or the right of the zero-correlation value by an ${\cal O}(1)$ factor, quantitatively changing the constraint for $f_a$. We can understand this behaviour from analytic estimations of the propagation of the variance for a multivariable function with correlations that we review in Appendix~\ref{app:corralation}. 
In this case, the variance of $\beta$ is\footnote{Note that, as discussed in  Appendix~\ref{app:corralation}, introducing a random sign, \jo{the  variance is given by Eq.~\eqref{eq:varwithsgn} , i.e. Var$[sgn(z)\phi_{\rm in}/f_a]=E^2(\phi_{\rm in}/f_a)+{\rm Var}[\phi_{\rm in}/f_a]$. The contribution of the random sign is encoded in the first term of Eq.~\eqref{eq:analyticvar}.}}:
\begin{equation}\label{eq:analyticvar}
    \sigma_\beta^2\simeq 0.033^2N\left(\frac{\langle{\phi}_i\rangle}{\langle{f_a}\rangle}\right)^2\left[1+\frac{\sigma_{f_a}^2}{\langle{f_a}\rangle^2}+\frac{\sigma_{\phi_i}^2}{\langle{\phi_i}\rangle^2}-4\rho\frac{\sigma_{\phi_i}\,\sigma_{f_a}}{\langle{\phi}_i\rangle \langle {f_a}\rangle}\right] \, ,
\end{equation}
where we neglected terms proportional to $\rho^2$ as well as higher-order correlators\footnote{
In our case, $\sigma_{f_a} / {\langle{f_a}\rangle}\simeq0.6$ and ${\sigma_\phi} / {\langle{\phi}\rangle}\simeq0.7$.}.
It is clear that a positive correlation requires a greater number of axions in order to give $\sigma_ \beta\sim 0.3$ deg compared to the uncorrelated case, which in turn gives a stronger upper bound on $f_a$. The opposite is true in the case of negative correlation. These results are summarized in Table~\ref{tab:tableQuadratic}.
Note that the previous expansion is not accurate when the distribution has long tails, since one should include more terms in Eq.~\eqref{eq:analyticvar}, thus the actual shape of the distribution matters and the result is not solely determined by the means and variances of the field amplitude and axion decay constant. In particular, we found that a broader spread of $\beta$ leads to an overall shift of the expected $N$ to lower values. This suppression relaxes the upper bound on the decay constant.

\begin{figure}
    \centering
    \begin{minipage}[b]{0.5\linewidth}
    \includegraphics[width=8cm]{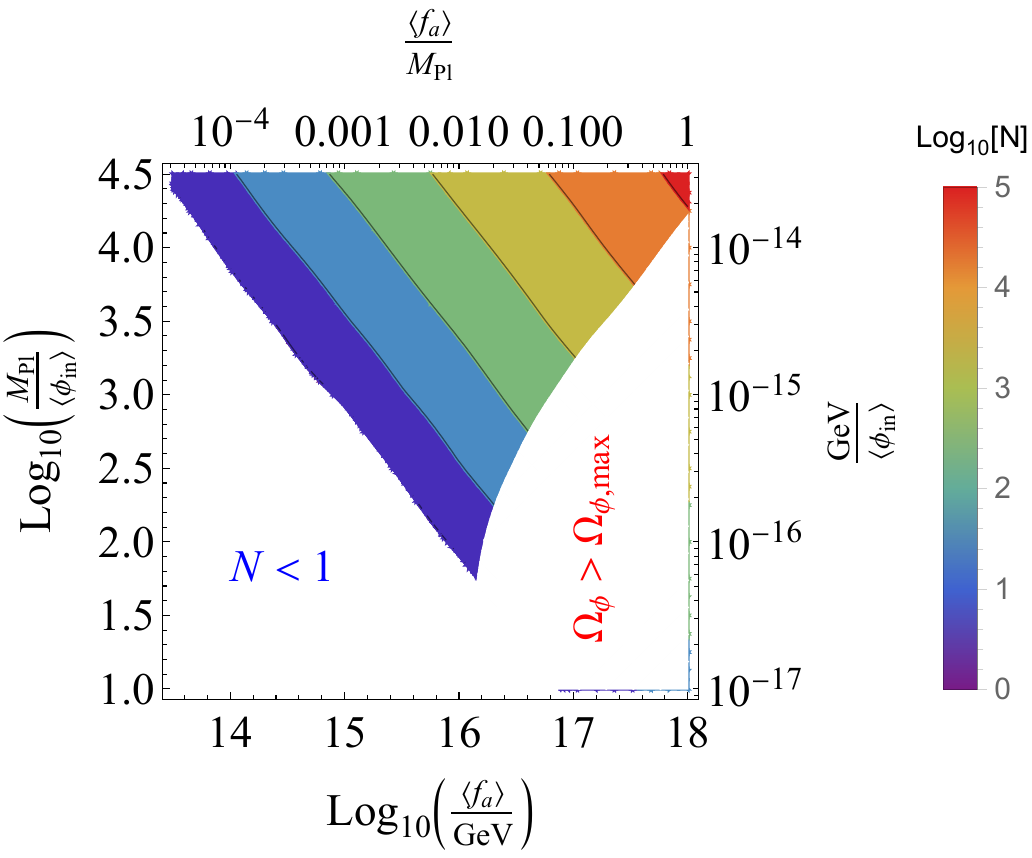}
    \end{minipage}
    \quad
    \begin{minipage}[b]{0.45\linewidth}
    \includegraphics[width=8cm]{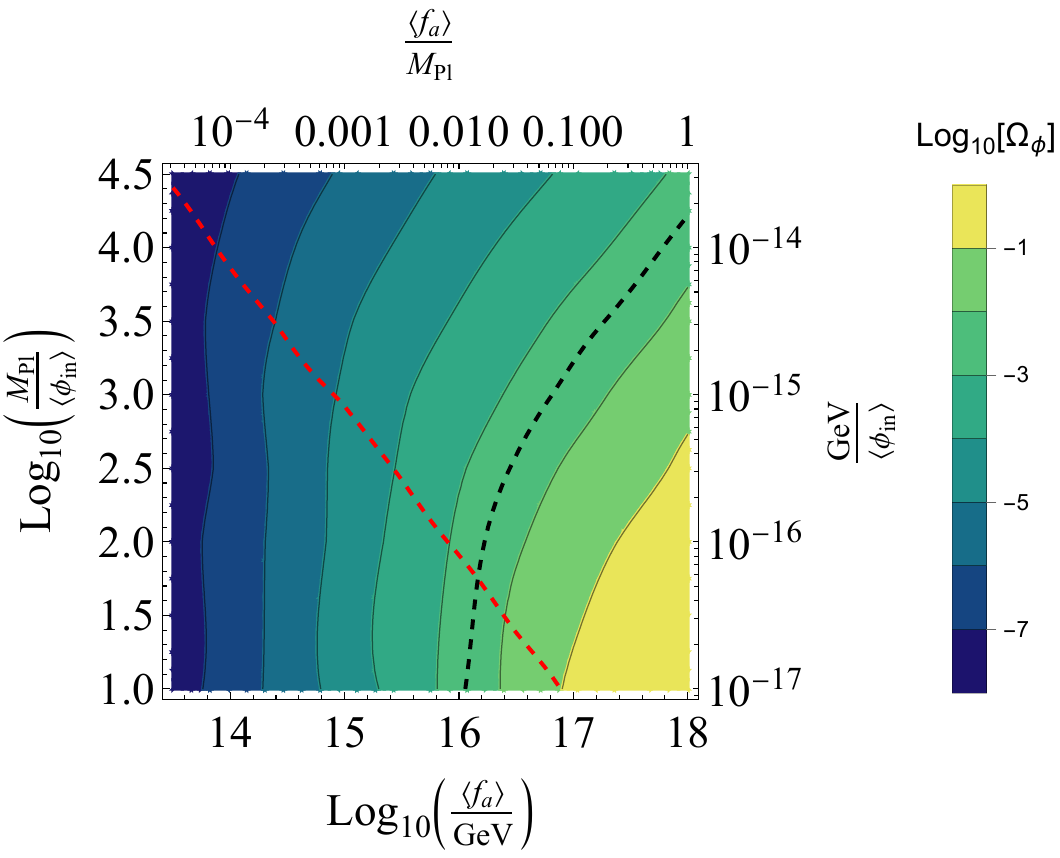}
    \end{minipage}
    \caption{\small \textit{Left:} Similar to Figure~\ref{fig:zerocorr}, but for the ``aligned'' case with $\rho=-0.8$. Compared to the previous scenarios, fewer axions are needed and their dependence is very different with respect to $\mu_\phi$ and $\mu_a$. \textit{Right:} The corresponding total axion abundance.}
    \label{fig:Alignedcase}
\end{figure}

We can compare the previous results to the case of ``aligned'' axions where $\beta$ scales linearly with the number of axions, thus a smaller value of $N$ is needed for given initial field values and decay constant, as shown in Figure~\ref{fig:Alignedcase}. In this case, the lines of constant $\Omega_\phi$  depend non-trivially on the initial value $\phi_{\rm in}$ and so does the maximum value of $f_a$. \jo{The numerical details of the ``aligned'' case are also described in Appendix~\ref{app:numdet}.}\\

\begin{table}[h!]
\centering
\begin{tabular}{|p{4cm}|p{4cm}|p{4cm}|}
\hline
      ${\rho(\phi_{\rm in},f_a)=0}$    & ${\rho(\phi_{\rm in},f_a)={\rm \jo{0.9}}}$  &  ${\rho(\phi_{\rm in},f_a)={{\rm \jo{-0.9}}}}$ \\
  \hline\hline
  $\langle f_{a}\rangle\lesssim6\times10^{16} {\rm GeV}$& $\langle f_{a}\rangle\lesssim2\times10^{16} {\rm GeV}$  & $\langle f_{a}\rangle\lesssim8\times10^{16} {\rm GeV}$\\
   $\langle \phi_{\rm in, max}\rangle\lesssim10^{17} {\rm GeV}$                        & $ \langle\phi_{\rm in, max}\rangle\lesssim8\times10^{16} {\rm GeV}$  & $\langle\phi_{\rm in, max}\rangle\lesssim10^{17} {\rm GeV}$ \\
  \hline
\end{tabular}
\caption{\small Summary of the constraints on the initial field value and the decay constant for the quadratic case for the different cases analysed in Figures~\ref{fig:zerocorr},\ref{fig:negcorr},\ref{fig:poscorr}}
\label{tab:tableQuadratic}
\end{table}

\subsection{Birefringence tomography}
\label{sec:quadratictomography}

\begin{figure}
    \centering
    \includegraphics[width=0.48\textwidth]{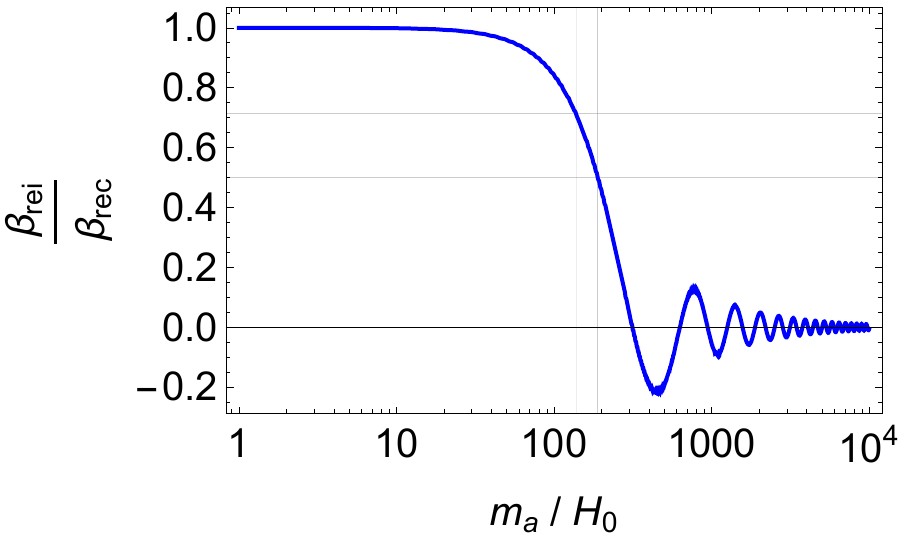}
    \includegraphics[width=0.45\textwidth]{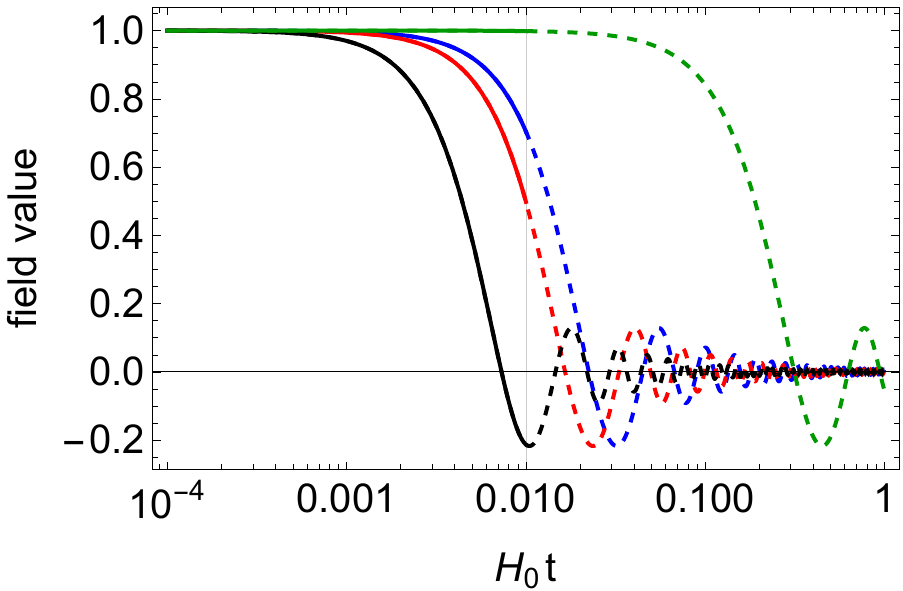}
    \caption{{\it Left:} The ratio $\beta_{\rm rei}/\beta_{\rm rec}$ for a single rolling axion as a function of its mass, normalized by $H_0$.
    {\it Right:} Four examples of rolling axions with $m/H_0=10, 140, 190, 430 $  (green, blue, red and black respectively). The rolling  before and after reionization is shown in solid and dashed respectively. \jo{The units for the field value are arbitrary, since the evolution does not depend on the initial field amplitude for a quadratic potential.}}
    \label{fig:betarecrei}
\end{figure}

\begin{figure}
    \centering
    \begin{minipage}[b]{0.5\linewidth}
    \includegraphics[width=8cm]{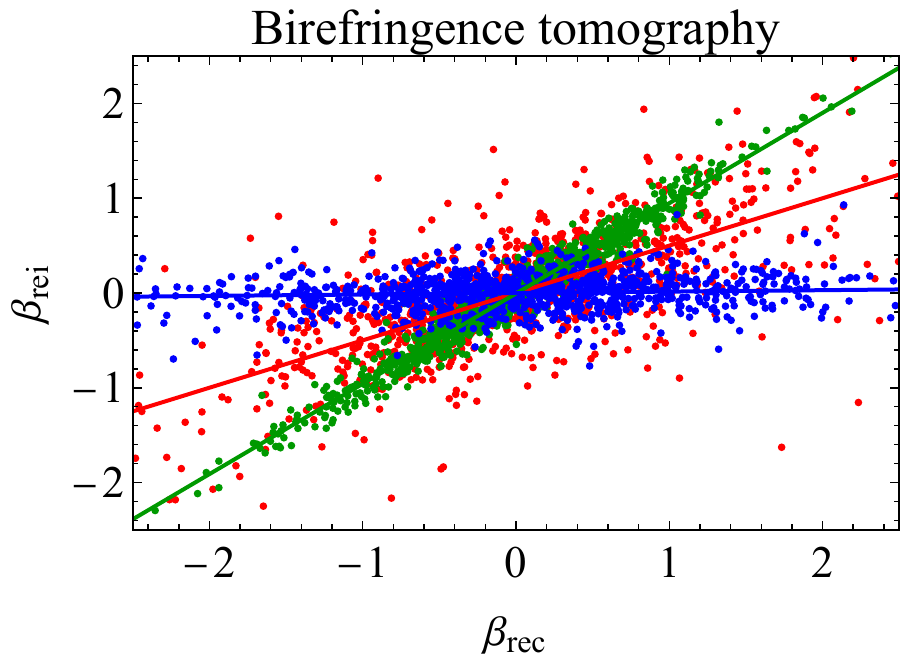}
    \end{minipage}
    \quad
    \begin{minipage}[b]{0.45\linewidth}
    \includegraphics[width=8cm]{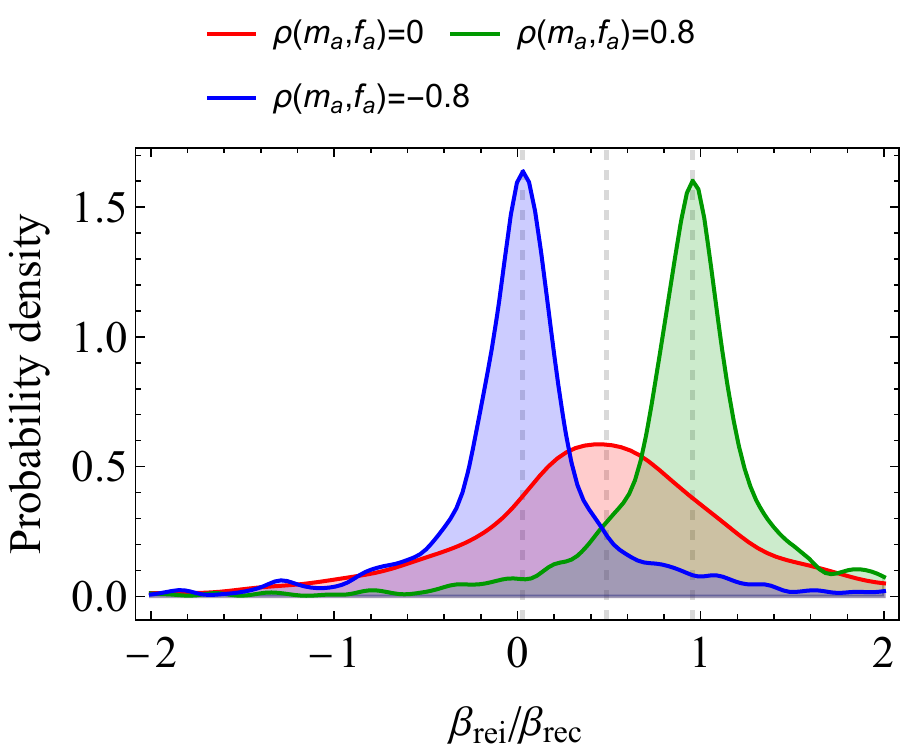}
    \end{minipage}
    \caption{\small \textit{Left:} Realizations of $\beta_{\rm rec}$ and  $\beta_{\rm rei}$ for three different cases of correlation $\rho(m,f_a)=0, 0.8,-0.8$. The points are scattered around the lines $\beta_{\rm rei}/\beta_{\rm rec}=0.5,0.95,0.01$ respectively.
   \textit{Right:} The PDF of $\beta_{\rm rei}/\beta_{\rm rec}$ for the three different cases.  } 
    \label{fig:PDFratio}
\end{figure}

\begin{figure}
    \centering
    \begin{minipage}[b]{0.5\linewidth}
    \includegraphics[width=7cm]{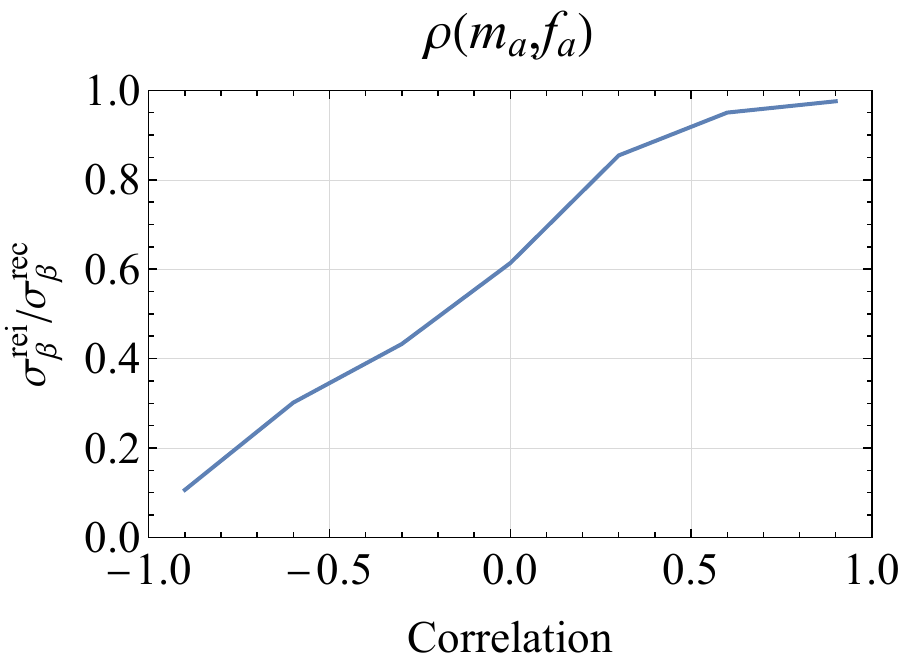}
    \end{minipage}
    \quad
    \begin{minipage}[b]{0.45\linewidth}
    \includegraphics[width=7cm]{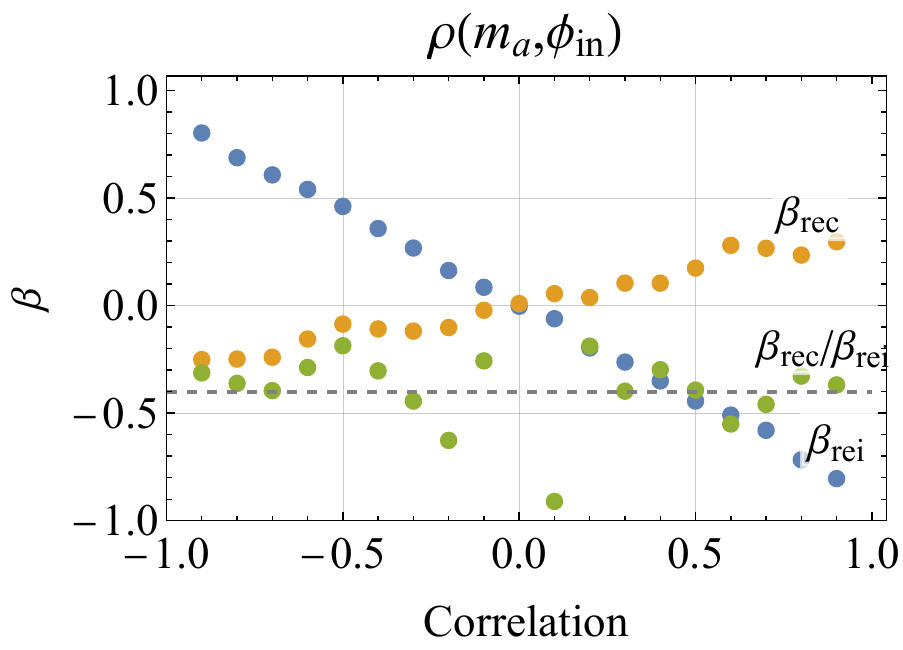}
    \end{minipage}
    \caption{\small \textit{Left:} The monotonic dependence of \jo{$\sigma^{\rm rei}_\beta/\sigma^{\rm rec}_\beta$ } on the correlation between the mass and the decay constant. \textit{Right:} \jo{We show the mean value $\langle\beta_{\text{rei}}\rangle$ and $\langle\beta_{\text{rec}}\rangle$ as a function of the correlation between the mass and the initial value. Contrary to the left panel the angle from reionization is bigger in amplitude than that from recombination, therefore, we take ${\langle\beta_{\text{rec}}\rangle}/{\langle\beta_{\text{rei}}\rangle}$. At large correlation, this oscillates around $-0.4$ as estimated by Eq.~\eqref{eq:weirdvalue}.} This is however only valid when the ratio is dominated by the mean of $\beta$ and not by its variance, as is the case close to $\rho=0$. } 
    \label{fig:Bratiocorr}
\end{figure}

In this section we discuss the expectations for the birefringence tomography \cite{Nakatsuka:2022epj,Galaverni:2023zhv}, i.e. the ratio of birefringence angles coming from reionization and recombination $\beta_{\text{rei}}/\beta_{\text{rec}}$, in the Axiverse scenario and  compare it to the single-field case. In the single-field  case, we can easily estimate this quantity using the analytical solution for the axion field evolution in the matter domination (MD) epoch $\phi(t)\propto \sin(mt)/mt$ \cite{Marsh:2015xka}, which gives
\begin{equation}
    \frac{\beta_{\text{rei}}}{\beta_{\text{rec}}}=\frac{\phi_{\text{rei}}-\phi_0}{\phi_{\text{rec}}-\phi_0}\simeq \frac{\sin(mt_{\text{rei}})}{\sin(mt_{\text{rec}})}\frac{t_{\text{rec}}}{t_{\text{rei}}} \, .
\end{equation}
This result is shown in Figure~\ref{fig:betarecrei} as a function of the axion mass\footnote{Here we take $\frac{t_{\text{rei}}}{t_0}=\Big(\frac{a_{\text{rei}}}{a_0}\Big)^{3/2}=(1+z_{\text{rei}})^{-3/2}\simeq 10^{-2}$ for $z_{\rm rei}\sim8$ and $\frac{t_{\text{rec}}}{t_0}\simeq10^{-4}$ for $z_{\rm rec}\sim 1000$.}. We also show the field evolution for four different axion masses. Notice that the dependence on the mass is rather strong when $m\sim H_{\rm rei}\sim 20 H_0$, which is why future detectors can use the difference in the birefringence between recombination and reionization to probe axions in the window $10\lesssim m/H_0\lesssim100$ \cite{Nakatsuka:2022epj}. 
\begin{figure}
    \centering
    \begin{minipage}[b]{0.5\linewidth}
    \includegraphics[width=8cm]{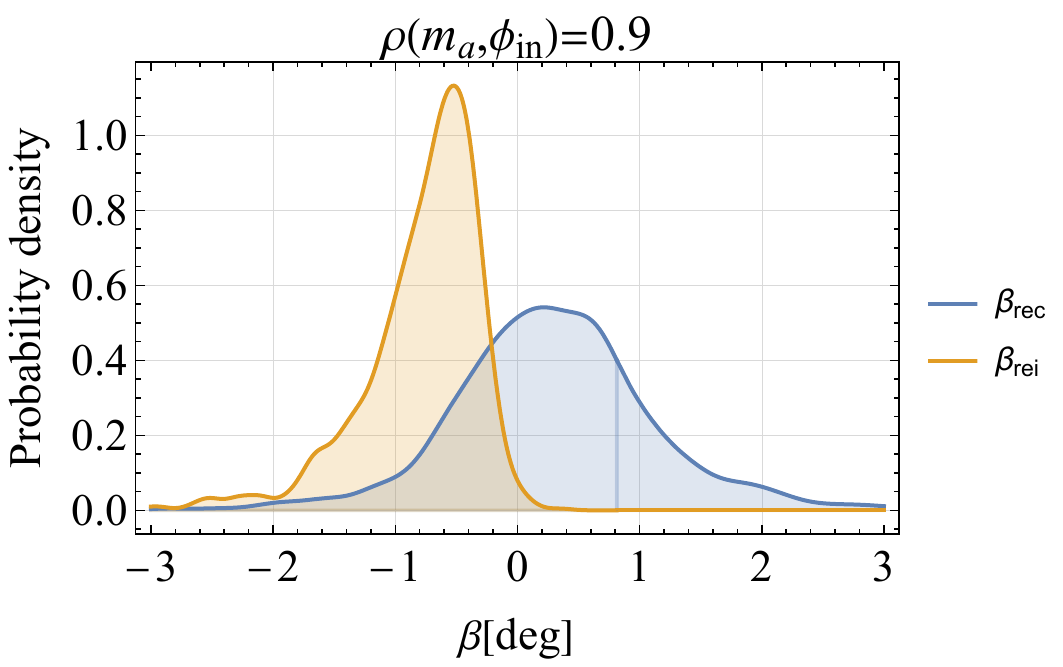}
    \end{minipage}
    \quad
    \begin{minipage}[b]{0.45\linewidth}
    \includegraphics[width=8cm]{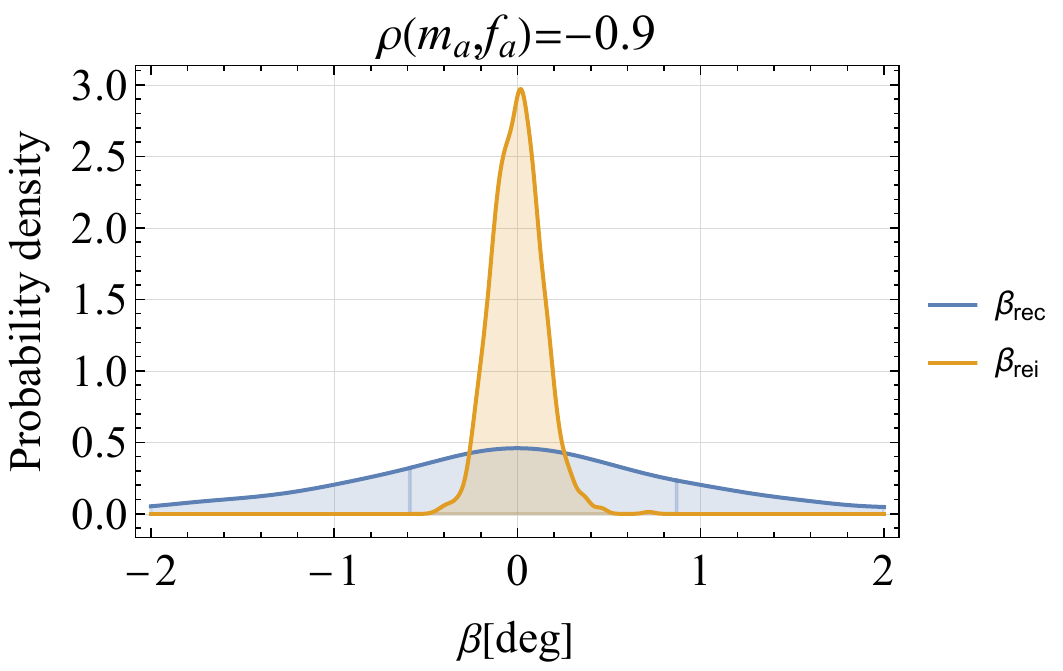}
    \end{minipage}
    \caption{\small The PDF of the birefringence angle from recombination and reionization in the presence of different types of correlations. \textit{Left:} An example of a large positive correlation between the mass and the initial field value. This leads to $|\langle \beta_{\rm rei}\rangle|> |\langle \beta_{\rm rec}\rangle|$  because of the partial cancellation of the birefringence from recombination from axions with different masses. \textit{Right:} An example of a large negative correlation between the mass and the decay constant (with a log-normal distribution). This mainly changes the variance of the distribution and not its mean value as in the previous case. }
    \label{fig:PDFrecrei}
\end{figure}
In the multiple-axion case, we can write 
\begin{equation}\label{eq:TOMO}
\frac{\beta_{\text{rei}}}{\beta_{\text{rec}}} = \frac{\sum^N_i(\phi_{\text{rei},i}-\phi_{0,i})/f_{a,i}}{\sum^N_i(\phi_{\text{rec},i}-\phi_{0,i})/f_{a,i}} \simeq \frac{\sum^N_i\frac{\phi_{\text{in},i}}{f_{a,i}}\left(\frac{\sin{(m_it_{\text{rei}})}}{m_it_{\text{rei}}}-\frac{\sin{(m_it_{0})}}{m_it_{0}}\right)}{\sum^N_i\frac{\phi_{\text{in},i}}{f_{a,i}}\left(1-\frac{\sin{(m_it_{0})}}{m_it_{0}}\right)}
\, ,
\end{equation}
where in the regime of interest $\phi_{\rm rec}\simeq\phi_{\rm in}$ \jo{and  we absorbed again the anomaly coefficient into $f_{a,i}$}. To make contact with the result given in Section~\ref{sec:cosinetomography}, notice that the contribution of axions with $mt_{\text{rei}}\gg1$ is suppressed, simply meaning that the axions that have started oscillating before reionization do not contribute to $\beta_{\text{rei}}$. From this and the fact that the average value of the birefringence is zero, we conclude that $\beta_{\text{rei}}/\beta_{\text{rec}}\sim \sigma^{\rm rei}_\beta/\sigma^{\rm rec}_\beta = 1/\sqrt{2}\simeq 0.7$ when the axions have a uniform mass distribution. This result can change significantly in the presence of correlations. For example, a correlation between the mass and the decay constant, \jo{ that we discussed in Section\ref{sec:statisticsALPs}\footnote{Note that in Appendix B of \cite{Mehta:2021pwf} a mild correlation of $\rho(m,f_a)\simeq0.5$ is discussed for string axions.}}, will weigh differently the contribution of axions rolling before and after reionization, enhancing the signal in one of the two regimes. As an example, for a strong negative correlation between the mass and the decay constant, the contribution of lighter axions is suppressed, which changes the variance of the distributions of $\beta_{\text{rec}}$ and $\beta_{\text{rei}}$. This is shown in Figure \ref{fig:PDFrecrei} for $\rho(m,f_a)=-0.9$, which leads to $\sigma^{\rm rei}_\beta/\sigma^{\rm rec}_\beta\simeq 0.1$, smaller than what we expect for zero correlation. On the other hand, for a positive correlation, the contribution from more massive axions is suppressed leading to $\beta_{\text{rei}}/\beta_{\text{rec}}\simeq1$ because most of the birefringence is generated at later times. Notice that even with this type of correlation, the birefringence ratio  is typically between zero and one, similar to the single-field case, as can be seen from the probability density functions shown in Figure~\ref{fig:PDFratio} for three different choices of $\rho(m,f_a)$ and from the monotonic dependence of its mean with the increase of the correlation shown in Figure~\ref{fig:Bratiocorr}.

Interestingly, in the multi-field case, it is also possible that $|\beta_{\text{rei}}/\beta_{\text{rec}}|>1$, but this requires  a cancellation between fields that roll after and before reionization. This can happen in the ``aligned" scenario with a correlation between the initial field value, randomly distributed around zero, and the mass. In this case, a positive correlation implies that more massive axions have preferably positive initial values whereas lighter ones have negative initial values and the opposite for a negative correlation. Because of this cancellation between positive and negative displacements, the mean values of the distributions of $\beta_{\text{rei}}$ and $\beta_{\text{rec}}$ change as  shown in the left panel of Figure~\ref{fig:PDFrecrei}. The shift of the mean values of the distribution of $\beta_{\rm rec}$ and $\beta_{\rm rei}$ are related to each other as can be intuitively expected. Indeed, approximating the distribution of $ \phi_{\rm in}$ and $m$ with a Binormal with $\rho\neq0$, the average value of the birefringence gets shifted proportionally to the correlation
\begin{equation}
    \langle \beta\rangle\propto\rho\sigma_\phi \sigma_{m}\frac{\partial }{\partial m}\frac{\sin{(mt)}}{mt}\Big|_{\langle m\rangle}.
\end{equation}
Inserting this into Eq.~\eqref{eq:TOMO}, it is straightforward to see that the ratio of the birefringence angles does not depend on the correlation \jo{
\begin{equation}\label{eq:weirdvalue}
    \frac{\langle\beta_{\text{rec}}\rangle}{\langle\beta_{\text{rei}}\rangle}= \frac{\cos{(\langle m\rangle t_{\rm rec})}-\cos{(\langle m\rangle t_{0})}}{\cos{(\langle m\rangle t_{\rm rei})}-\cos{(\langle m\rangle t_{0})}}\sim \frac{1-\cos(100)}{\cos(1)-\cos(100)}\sim -0.4, 
\end{equation}
where} $\langle m\rangle\simeq 10^2H_0$ for uniform distributions of masses within $H_0<m<10^4H_0$. This simple result is surprisingly accurate when compared to the numerical calculation of ${\langle\beta_{\text{rec}}\rangle}/{\langle\beta_{\text{rei}}\rangle}$ as shown in Figure~\ref{fig:Bratiocorr} for sufficiently large correlation, when the ratio is dominated by the mean and not the variance; this condition breaks down for $\rho\sim0$. \jo{Note that in Figure~\ref{fig:Bratiocorr} we take ${\langle\beta_{\text{rec}}\rangle}/{\langle\beta_{\text{rei}}\rangle}$ and not the inverse because for ${\langle\beta_{\text{rec}}\rangle}<{\langle\beta_{\text{rei}}\rangle}$ the numerical error is smaller.}
It is worth noting that a similar situation was recently discussed in Ref.~\cite{Namikawa:2023zux} for a two-axion model to motivate the search of birefringence originated at late-time through the recently proposed polarized Sunyaev Zel’dovich effect \cite{Hotinli:2022wbk,Lee:2022udm}. This search is further motivated in our axiverse scenario where the signal is generated at all cosmic times. 

\section{Axion Monodromy}
\label{sec:monodromy}

We now move to the more complicated case of axion monodromy potentials  \cite{McAllister2008, Silverstein,Kaloper:2008fb,Flauger,Kaloper:2011jz,Kaloper:2014zba,Kaloper:2016fbr,DAmico:2021vka,DAmico:2016jbm}. 
Our goal is to explore the parameter space of the Axiverse for this type of potential since altering the large-field dynamics of the rolling axion field can affect both the birefringence \jo{angle} as well as the axion abundance. 
We model axion monodromy as
\beq
\label{eq:monodromy}
V(\phi) = \frac{M^2 m^2}{2p} \left [
\left(\frac{\phi^2}{ M^2}+1\right )^p-1
\right ]
\eeq
where $p$ is a free parameter that controls the large-amplitude slope of the potential. The value $p=1/2$ leads to the usual asymptotically linear potential, but values of $p<1/2$ have appeared in the literature in the context of inflation \cite{McAllister:2014mpa}. 
Furthermore, the mass-scale $M$ defines the transition from a quadratic potential for $|\phi|<M$ to a potential that grows like $|\phi|^{\,2p}$ for $|\phi|>M$. Choosing $M$ close to $\phi_{\rm in}$ will largely reproduce the results shown in the previous section, as the dynamics will be dominated by the quadratic part of the potential. The pre-factor of the potential is chosen, such that $m$ is the free-particle mass of the axion.
Axion monodromy potentials of the form of Eq.~\eqref{eq:monodromy} introduce two new parameters, compared to the quadratic case, the tranisiton scale $M$ and the power-law exponent $p$.

We start by examining the axion abundance $\Omega_\phi$ as a function of $M$, $\phi_{\rm in}$ and $p$. 
The potential of Eq.~\eqref{eq:monodromy} can be approximated as:
\begin{equation}
		V(\phi) = 
		\begin{cases} 
		\frac{M^2 m^2}{2p}
\left(\frac{|\phi|}{ M}\right )^{2p} & \text{if } |\phi|>M \\
		\frac{1}{2}m^2\phi^2  & \text{if } |\phi|<M. 
		\end{cases}
\end{equation}
Because at large field values, the potential is flatter than in the quadratic case, the onset of  oscillations is delayed. This can be estimated as in Ref.~\cite{Kitajima:2018zco}, by comparing the two relevant time-scales, that of  cosmic expansion $t_H\simeq H^{-1}$ and that of the motion driven by the potential $t_\phi=\omega_\phi^{-1}$.
Defining
$\omega_\phi\simeq\sqrt{\left|{V_{,\phi}}/{\phi}\right|}$ leads to
$H_{osc}\simeq3\omega_\phi$ and, by using the scaling $a(t)\propto t^{2/3}$ in the MD era, we find that the field starts rolling, and subsequently oscillating, around:
\begin{equation}\label{eq:tauosc1}
    t_{osc}=2\sqrt{\left|\frac{\phi}{V_{,\phi}}\right|}=\left(\frac{2}{m}\right)\left(\frac{\Tilde{\phi}}{M}\right)^{1-p} \, ,
\end{equation}
where $\tilde{\phi}
= M (H/3m)^{1\over p-1}$ is the field value evaluated at that time.
For the remainder of this section, we work with non-dimensional units $t\to H_0 t$, $ m\to m/H_0$ and $\phi\rightarrow\phi/M$, unless otherwise stated.

It is important to note that the parameter space of interest for having cosmic birefringence, i.e. the field rolls significantly between recombination and today $10^4<t_{osc}<t_0$, does not depend only on the axion mass, as is the case for a quadratic potential, but is very sensitive to the initial field value and the power-law parameter $p$. In particular, the maximum axion mass that can contribute to the signal is lifted to higher values as $m_{\text{max}}\simeq 10^4(\phi_{\rm in})^{1-p}\gg10^4$ (in units of $H_0$) thus opening the available parameter space to masses $m > 10^{4}$. The cosmological evolution of the background field in these potentials is more complicated since  at large field amplitude $|\phi|\gg1$,  the abundance redshifts as \cite{Turner:1983he}
\begin{equation}\label{eq:ABvsp}
    \rho\propto a^{-3(1+\omega)} \qquad \text{with}\qquad w=\frac{p-1}{p+1},
\end{equation}
and later when the field only explores the quadratic region around the minimum, it dilutes as standard dark matter with $w=0$. In what follows we focus on the usual $p=1/2$ case with linear behaviour at large field values. We study in detail how the background field redshifts and discuss how the phenomenological constraints from birefringence change in this case.

\subsection{
Asymptotically linear potential}\label{sec:monodromy1/2}

We now focus on the case of $p=1/2$, where the potential can be approximated as:
\begin{equation}
		V(\phi) = 
		\begin{cases} 
		{M^2 m^2}
\left(\frac{|\phi|}{ M}-\frac{1}{2}\right ) & \text{if } \frac{|\phi|}{ M}>1\\
		\frac{1}{2}M^2m^2\left(\frac{\phi}{ M}\right)^2  & \text{if } \frac{|\phi|}{ M}<1 
		\end{cases}
\end{equation}
and we briefly reintroduced the mass scale $M$ for clarity. 
Taking the initial condition as $\phi_{\rm in}\gg1$, the axion field starts evolving in the linear potential where the analytic solution is \cite{Weinbergcos}
\begin{equation}\label{eq:linearev}
    \phi_{\rm lin}(t)=\phi_{\rm in}-m^2t^2/6 \, .
\end{equation}
After the field reaches unity $\phi=1$ in a time-scale of 
\beq\label{eq:ttr}
t_{tr}=\sqrt{6(\phi_{\rm in}-1)}/m \, ,
\eeq  a first ``transition'' of dynamical behaviour occurs. The subsequent evolution, in particular the damping of the oscillations, depends on the value of $\phi_{osc}$, which defines when the potential term becomes dominant compared to the cosmological term $3H(t)$, which acts as a time-dependent viscosity.

When $\phi_{\rm in}\gg1$ 
the field abundance redshifts as in Eq.~\eqref{eq:ABvsp} with $w=-1/3$, until the field amplitude is sufficiently damped ($\phi<1$) and the abundance redshifts as $w=0$. Therefore, depending on the initial condition, we can distinguish three types of behaviour:  rolling in a linear potential, oscillating and probing the linear part of the potential and finally oscillating in a quadratic potential. \jo{We identify the transition between oscillation in linear and quadratic potential with the time $t_{1}$ corresponding to field amplitude of $\phi=1$ as discussed later. }
In what follows we identify the relative parameter space for each behaviour to occur: 

\begin{description}
    \item[(i) Oscillations in the quadratic potential $t_{tr}>t_{osc}$]    
  When the initial value is smaller than unity from the beginning, the field never probes the linear part of the monodromy potential. \jo{Actually, this regime extends until $\phi_{\rm in}<3M$ because at the time the field starts oscillating $\phi<M$ and the field evolution continues as in the quadratic case. This can be seen from the linear evolution given in Eq.~\eqref{eq:linearev} evaluated at the time when the oscillations start \eqref{eq:tauosc1}:
    \begin{equation}
        \phi_{osc}=\phi_{\rm in}-\frac{1}{6}m^2t_{osc}^2 = \phi_{\rm in}-\frac{1}{6}m^2\frac{4\phi_{\rm in}}{m^2} \, ,
    \end{equation}
  leading to $\phi_{osc}={\phi_{\rm in}}/{3}$.
 } The abundance is given by Eq.~\eqref{eq:Omegaquadest}, as in the quadratic case.
   
    \item[(ii) Oscillations in the linear potential and no transition $t_{osc}<t_{tr}<t_0<t_{1}$]
    This happens when $\phi_{\rm in}>1$, thus the evolution of the axion abundance includes the rolling in the linear part of the axion potential and the subsequent oscillations, which also probe the linear part of the potential. The evolution of the abundance during the first part of the motion can be accurately computed from the analytic solution of Eq.~\eqref{eq:linearev} for $t<t_{tr}$ and from Eq.~\eqref{eq:ABvsp} for $t>t_{tr}$.\footnote{The reason why $t_{tr}$ better approximates  the time in which the abundance passes from a linear rolling to oscillations compared to $t_{osc}$ is that $t_{osc}$ does not take into account the total time the field needs in order to reach the minimum of the potential and then start to oscillate.} 
    The final value of the abundance becomes (we neglect numbers smaller than $\mathcal{O}(1)$ compared to $\phi_{\rm in}$):
    \begin{equation}\label{eq:ABlinear}
        \Omega_{\phi}\simeq 0.36 \Big(\frac{M}{M_{\rm Pl}}\Big)^2\phi_{\rm in}^{5/3}m^{2/3}.
    \end{equation}

   During the evolution, the amplitude of the oscillations gets reduced and the field will eventually probe only the quadratic part of the potential. \jo{ Numerically we find that $\phi(t_{1})=1$ well approximates this transition. 
   For $p=1/2$ we find that the amplitude of the scalar field scales as the density from Eq.~\eqref{eq:ABvsp} $ \phi_{\rm amp}(t)\sim\phi_{osc}\left({t_{osc}}/{t}\right)^{4/3}$, which gives
\begin{equation}
\label{eq:t12}
        t_1=0.44\frac{\left(\phi_{\rm in}\right)^{5/4}}{m}\, .
    \end{equation}
    }

    Given the values of $m$ and $\phi_{\rm in}$, this transition hasn't happened yet if $t_{1}>t_0\sim 1$ (in  units of $H_0$). 

    \item[(iii) Oscillations in the linear part and transition to quadratic potential $t_{osc}<t_{tr}<t_{1}<t_0$]
    Following the previous discussion, in this regime, the evolution of the abundance has three different scalings: rolling and oscillations in the linear potential and subsequent oscillations in the quadratic region. In Figure \ref{fig:Backevol} we show that the analytical scaling in these three phases closely follows the numerical evolution and gives the current abundance\jo{ 
    \begin{equation}\label{eq:Ablinplustrans}
        \Omega_{\phi}\simeq 0.2 \Big(\frac{M}{M_{\rm Pl}}\Big)^2\phi_{\rm in}^{15/6}.
    \end{equation}
    }
    Remarkably, as for the quadratic potential case, the final abundance does not depend on the mass of the field.
\begin{figure}
    \centering
    \begin{minipage}[b]{0.5\linewidth}
    \includegraphics[width=5.5cm]{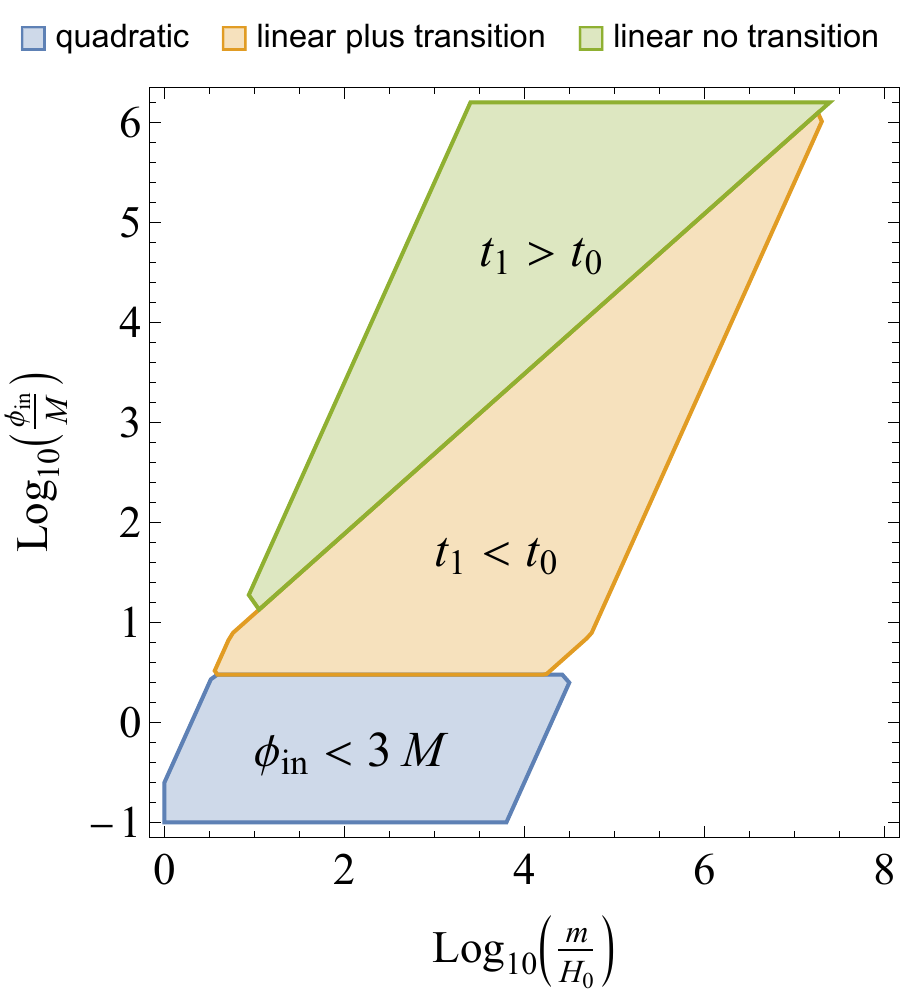}
    \end{minipage}
    \quad
    \begin{minipage}[b]{0.45\linewidth}
    \includegraphics[width=7.5cm]{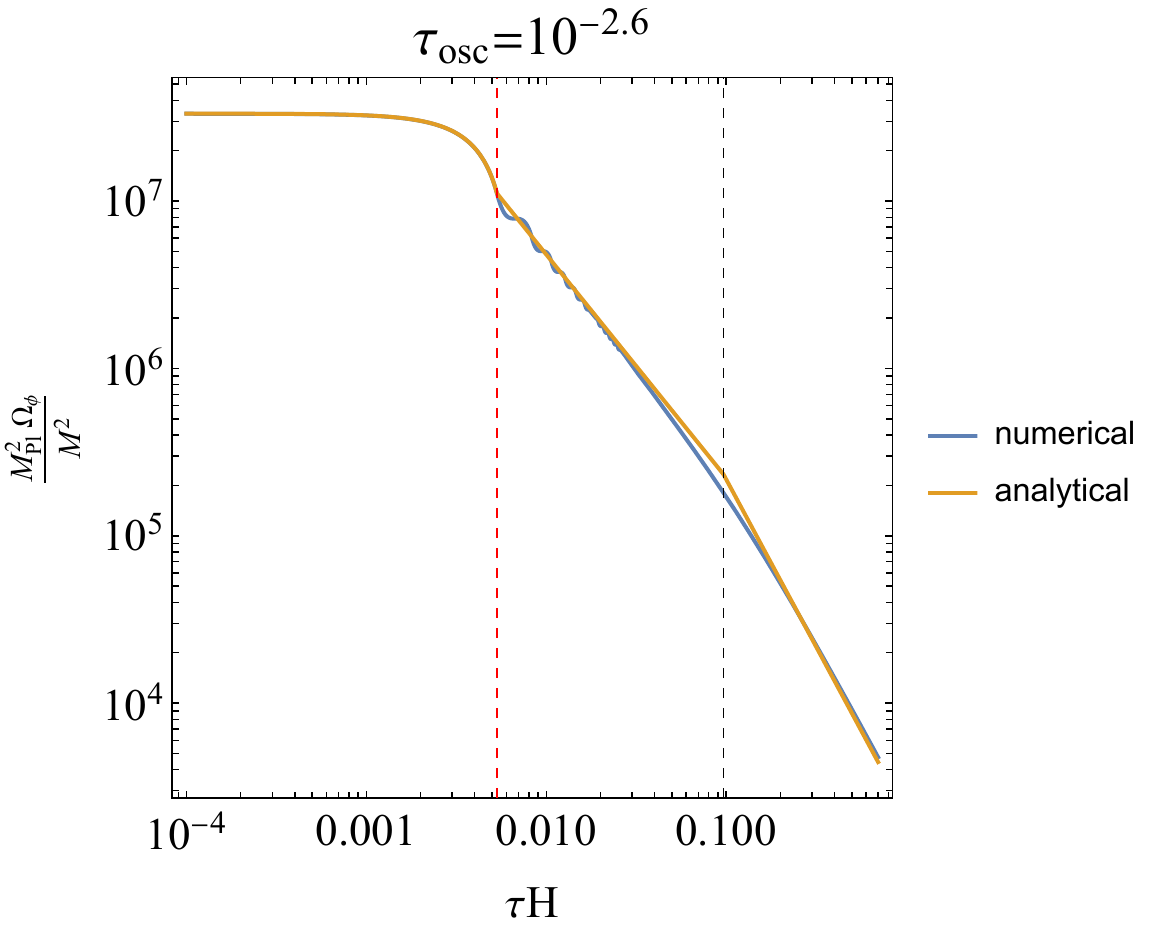}
    \end{minipage}
    \caption{\small \textit{Left:} Regions of the parameter space of the initial values and masses corresponding to the different evolution for the asymptotic linear potential explained in Section~\ref{sec:monodromy1/2}  with $10^{-4}<t_{osc}<1$. The green region corresponds to fields that oscillate exploring the large field values 
    and the transition from oscillation in linear to quadratic potential hasn't occurred yet.
    Fields in the orange region experience such a transition within the present time and are oscillating in the quadratic region around the minimum of the potential. Evolution in the blue region is quadratic \jo{corresponding to $\phi_{\rm in}\leq 3 M$}. 
    \textit{Right:} An example of the background evolution of an axion field that experiences a transition from the oscillations in the linear to the quadratic regime of the potential. The chosen values are $m=10^4$ and the time-scales $t_{osc}=10^{-2.6}$  and $t_{1}=0.1$ \jo{are denoted by the dashed red and black vertical lines respectively}. }
    \label{fig:Backevol}
\end{figure}
\end{description}

{The regions of parameter space corresponding to the three different types of evolution are shown in the left panel of Figure~\ref{fig:Backevol}. }

\subsubsection{Probability of rolling after recombination}

A natural question one can ask is: how likely is it that the fields start rolling after recombination, i.e.  $10^{-4}<t_{osc}<1$,  given an initial probability distribution of $\phi_{\rm in}$ and $m$ and how much does it change for the asymptotically linear potential?

In the case of a cosine or quadratic potential, the onset of oscillations depends solely on the mass (assuming that the initial field amplitude is not fine-tuned to be close to the maximum of the cosine potential), thus the number of active axions after recombination depends on how many populate the mass range $1\lesssim m/H_0\lesssim 10^4$.  If we take the mass uniformly distributed in log-space over several orders of magnitude, e.g. $\log_{10}(m/\text{eV})\in[-33\log_{10}(m_{\rm min}/H_0),27]$, this is given by $N_{\rm tot}\times 4/60$ \jo{for $m_{\rm min}=H_0$}.
For the monodromy case, Eq.~\eqref{eq:tauosc1} shows that the onset of oscillations depends also on the initial distribution of $\phi_{\rm in}$. In particular, the time of the onset of oscillations in logarithmic space is: 
\begin{equation}
    \log_{10}{t_{osc}}=\log_{10}(2) -\Big[33 + \log_{10}{\frac{m}{\text{eV}}}\Big]+ (1 - p)\Big[-\log_{10}{\frac{\phi_{\rm in}}{M_{\rm Pl}}} + 30\Big] \, ,
\end{equation}
thus its distribution is given by the convolution of the individual distributions of  $\log(\phi_{\rm in})$ and $
\log(m)$. When the initial field has a distribution peaked at a certain scale, we found that the probability of having $10^{-4}<t_{osc}<1$ does not change much compared to the previous case, since the mass is distributed over many orders of magnitude
${\cal O}\left ( {4}/(\log_{10}\left (m_{\rm max}/m_{\rm min}\right) \right )$. As can be seen from Figure \ref{fig:Backevol}, the masses  
of axions active between recombination and today are still around four orders of magnitude, but, compared to the quadratic case, the mean value gets shifted to higher masses as $\phi_{\rm in}/M$ increases. This means that the probability is not affected if the mass distribution is flat, but it is certainly \jo{altered} if the distribution is peaked at some scale or if it is tilted towards higher or lower masses.

\subsection{  Range of transition scale values for Monodromy potentials }
\label{sec:monodromyparameterspace}

The main goal of this work is to use the birefringence signal and the DM fraction to determine the interesting parameter space of $(\phi_{\rm in}, f_a, M, m)$ where the signal can be explained in the Axiverse scenario without conflicting with constraints on the current axion abundance. In Section~\ref{sec:monodromy1/2} we derived the relation between $(\phi_{\rm in},M,m)$ and the present axion abundance, which gives an upper bound on the transition scale in the different regimes. On the other hand, the birefringence angle connects the field amplitude to the decay constant $f_a$ \jo{and from the astrophysical bound on $f_a$ we can} infer a lower bound on $M$. \jo{To our knowledge, one of the tightest constraints on $f_a$ comes from the absence of evidence of spectral distortions of X-rays flux coming from quasars (e.g. Ref.~\cite{Reynes:2021bpe}) which leads to $f_a\gtrsim 10^{9}$ GeV.
This bound is derived assuming one axion and not multiple light axions, as in our case. In the latter case, the photons can convert to all axions lighter than $m_a\ll 10^{-12}$ eV~\cite{Reynes:2021bpe} which can be treated as massless. Following Ref.~\cite{Halverson:2019cmy}, we can define a single axion that couples to $F_{\mu\nu}$ and has an effective coupling $g^{\rm eff}_{\phi\gamma}=\sqrt{\sum_i g_{\phi\gamma,i} }$. If the coupling of different axions is uncorrelated and is distributed around $\langle g_{\phi\gamma}\rangle$, one naively expects that the coupling is enhanced by a factor $\sqrt{N_{12}}$ where $N_{12}$ is the fraction of axions with mass below the threshold mass $m_a\ll 10^{-12}$ eV\footnote{In~\cite{Reynes:2021bpe} it is shown that considering some string theory compactifications, the scaling can be stronger than $\sqrt{N}$.}. We confirm numerically that taking the distribution of $f_a$ given in Eq.~\eqref{eq:PDFs} the effective coupling scales as $\sqrt{N}$\footnote{This scaling is modified if one increases the variance of the log-normal distribution, because the coupling gets dominated by the minimum decay constant. Numerically, we find that for $\sigma_\phi>1$ the $\sqrt{N}$-scaling is not accurate at large N.}, therefore we take $g^{\rm eff}_{\phi\gamma}=\sqrt{N_{\rm dec}\times \log_{10}(m_{12}/m_{\rm min})}\alpha_{\rm EM}/2\pi\langle f_a\rangle$ where $m_{12}=10^{-12}$ eV and $m_{\rm min}=10^{-33}$ eV, thus $\log_{10}(m_{12}/m_{\rm min})=21$. With this, we proceed with the scan of allowed values for the scale of the transition in the various regimes:}

\begin{description}
    \item[(i) Quadratic regime \jo{$( \langle\phi_{\rm in}\rangle\ll M)$}]

    In this case, using Eq.~\eqref{eq:Omegaquadest} we get 
    \begin{equation}
    \Omega_\phi\simeq \frac{3}{8}\Big(\frac{M}{M_{\rm Pl}}\Big)^2 \langle\phi_{\rm in}^2\rangle < \frac{3}{8}\Big(\frac{M}{M_{\rm Pl}}\Big)^2< 10^{-2} \, , 
    \end{equation}    
    leading to the upper bound $ M<0.1M_{\rm Pl}$ which gets reduced by $1/\sqrt{4N_{\rm dec}}$ in the case of multiple axions.
    Inverting Eq.~\eqref{eq:analyticvar}  for the variance of $\beta$ leads to the scaling $\sqrt{4N_{\rm dec}} \langle\phi_{\rm in}\rangle\simeq 10 \langle f_a\rangle$, thus reintroducing the transition scale $ \langle\phi_{\rm in}\rangle\simeq 10 \langle f_a\rangle/\sqrt{4N_{\rm dec}}<M$ and the upper bound on the decay constant $\langle f_a\rangle/\sqrt{21N_{\rm dec}}\gtrsim 10^{9}$ GeV leads to
    \begin{equation}
        10^{10}\text{GeV}<M<0.1 M_{\rm Pl}/\sqrt{4N_{\rm dec}}.
    \end{equation}

    \item[(ii) Linear regime]
    In this regime, there is a minimum initial value shown as the bottom corner of the \jo{green} region of Figure \ref{fig:Backevol} that is \jo{$\langle\phi_{\rm in}\rangle\simeq10M$} for $m\sim H_0$, which gives the upper bound \jo{$M\lesssim0.01 M_{\rm Pl}$} from the abundance constraint. However, there is no minimum value for the transition scale that would come from a maximum value of $\langle\phi_{\rm in}\rangle$, as we can keep $t_{osc}$ constant when $\phi_{\rm in}$ increases by increasing $m$. Moreover, one can always adjust the transition scale to not exceed the abundance constraints,  $ {M^2}/{M_{\rm Pl}^2}\ll1$ as follows from Eq.~\eqref{eq:ABlinear}. 
    
    \item[(iii) Linear plus transition]
    The maximum transition scale for which the field starts in the linear regime is given by the abundance constraint when $\langle\phi_{\rm in}\rangle\sim3 M$, which gives $\sqrt{4N_{\rm dec}}M/M_{\rm Pl}\lesssim1$. On the other hand, the minimum transition scale can be related to the maximum initial value $\langle\phi_{\rm in}\rangle\lesssim 10^6 M$ (corresponding to the upper corner of the orange region in Figure \ref{fig:Backevol}) and the birefringence through 
    \begin{align}
        &\frac{\langle f_a\rangle}{M}\sim \frac{\sqrt{N}}{10}\frac{\langle\phi_{\rm in}\rangle}{M}<10^5 \sqrt{4N_{\rm dec}}  \, ,
        \end{align}
        leading to
     $  10^{4}\text{GeV}\lesssim M\lesssim M_{\rm Pl}/ \sqrt{4N_{\rm dec}}    $,
    where we used the lower bound for the decay constant. 
\end{description}

\subsection{Implications from Cosmic Birefringence}

\begin{figure}
    \centering
    \begin{minipage}[b]{0.45\linewidth}
    \includegraphics[width=7.5cm]{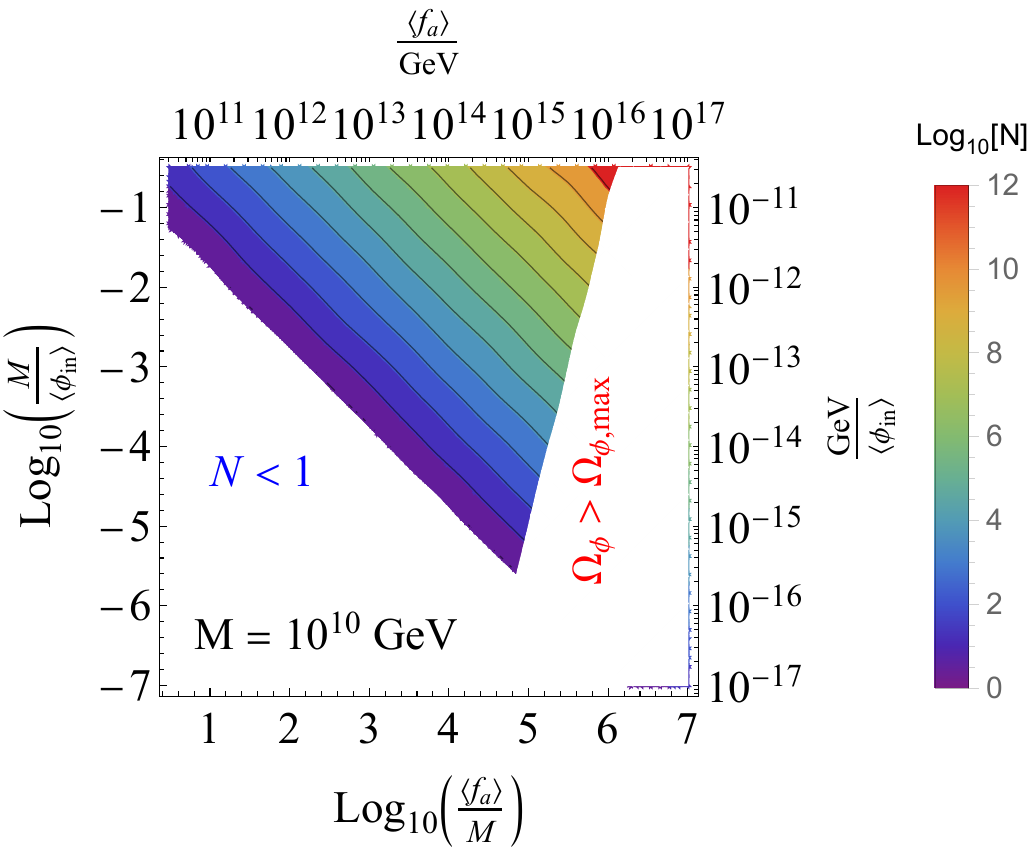}
    \end{minipage}
    \quad
    \begin{minipage}[b]{0.45\linewidth}
    \includegraphics[width=7.5cm]{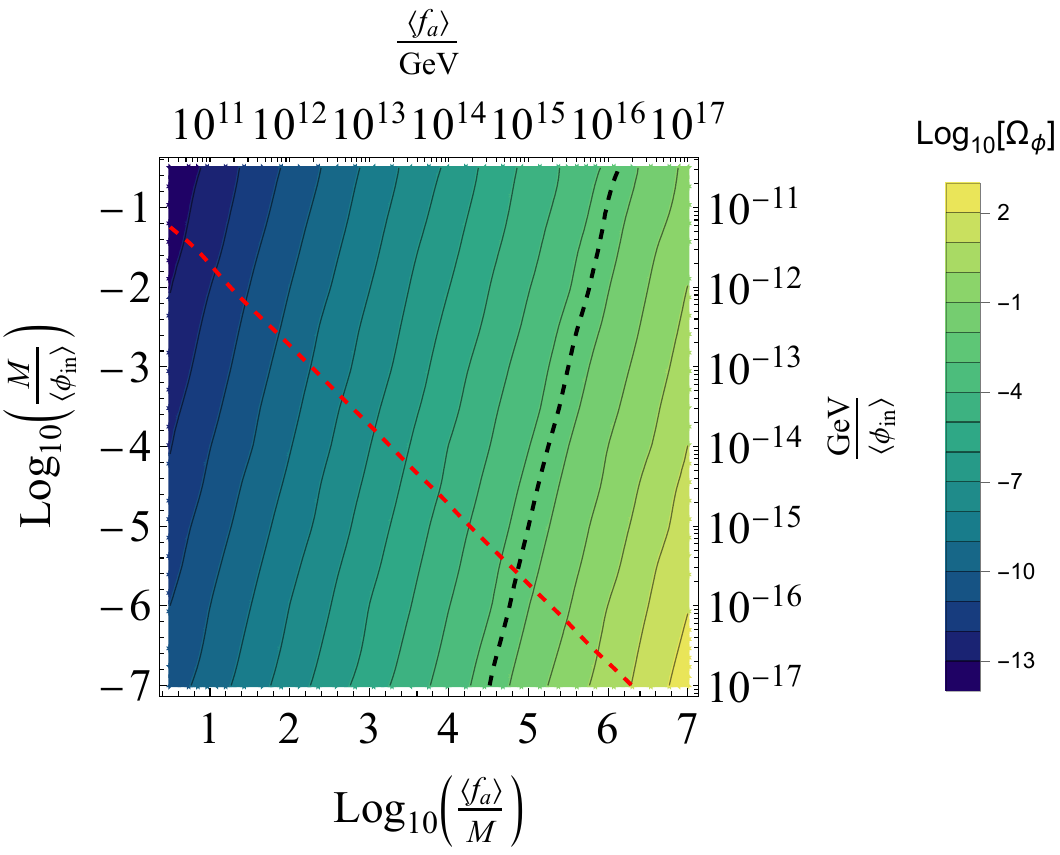}
    \end{minipage}
    \caption{\small \textit{Left:} Number of axions needed to achieve $\sigma_\beta=0.3$ deg as a function of $\mu_\phi$ and $\mu_a$ and corresponding excluded region for the ``linear plus transition" case, i.e. orange region in Figure \ref{fig:Backevol}. The transition scale is fixed to $M=10^{10}$ GeV. \jo{Compared to the quadratic case the maximum initial field value is lower by an order of magnitude $\langle\phi_{\rm in}\rangle\lesssim 10^{16}$ GeV. } \textit{Right:} Corresponding total abundance for the number of axions coming from the left panel. Note the negative correlation between the initial field value and decay constant of regions with constant abundance.  }
    \label{fig:10Gev LPTres}
\end{figure}
\begin{figure}
    \centering
    \begin{minipage}[b]{0.45\linewidth}
    \includegraphics[width=7.5cm]{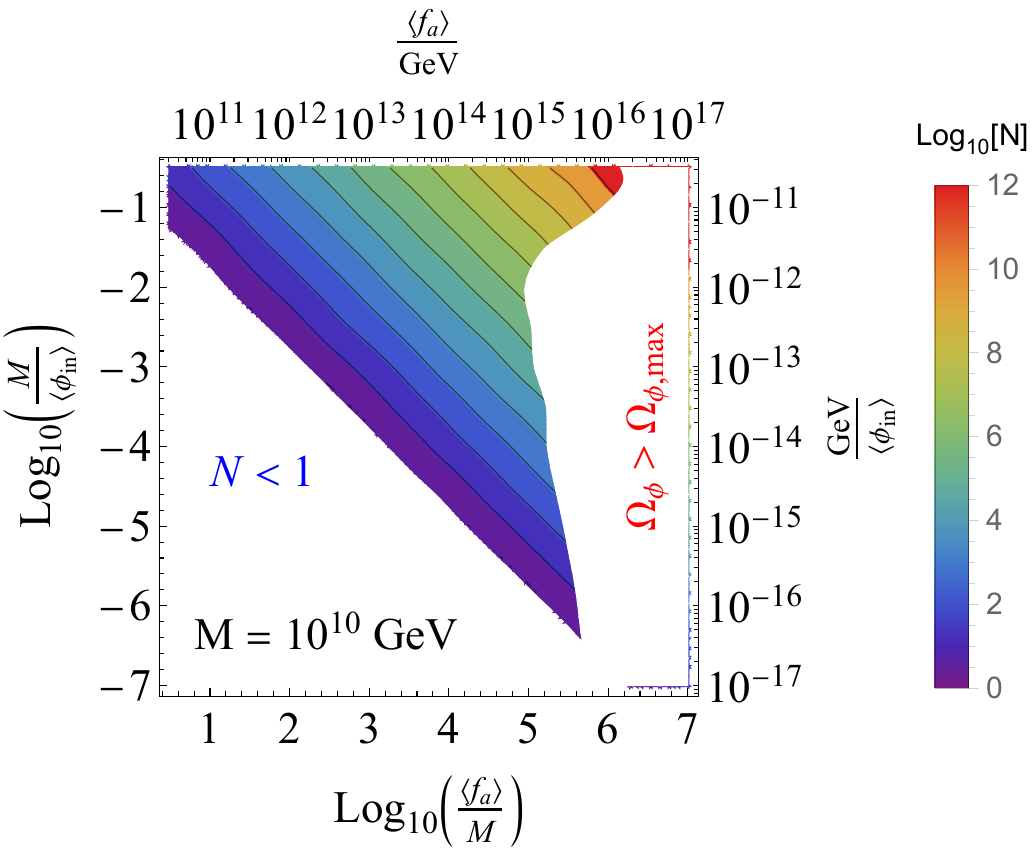}
    \end{minipage}
    \quad
    \begin{minipage}[b]{0.45\linewidth}
    \includegraphics[width=7.5cm]{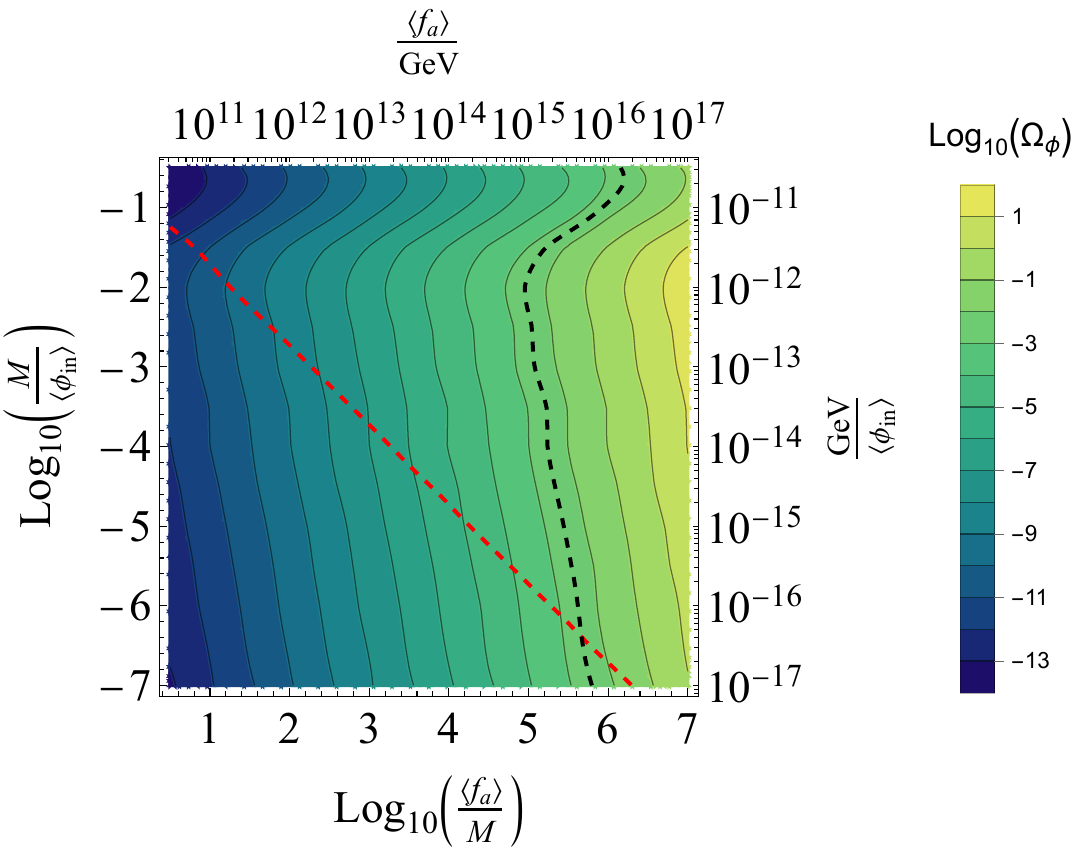}
    \end{minipage}
 \caption{\small Analogous to Figure \ref{fig:10Gev LPTres} for initial conditions leading to just linear evolution, i.e. the green region in  Figure~\ref{fig:Backevol}. Note that in this regime, the abundance contours present a change of behaviour, similar to a ``knee", because of the dependence of the mass on the formula of the final axion abundance in Eq.~\eqref{eq:ABmeanLin}. As the initial field value increases, the mean value of the axion mass contributing also increases, as shown in Figure~\ref{fig:parspacetrans}.  } \label{fig:10Gev linear}
\end{figure}

\begin{figure}
    \centering
    \includegraphics[scale=0.6]{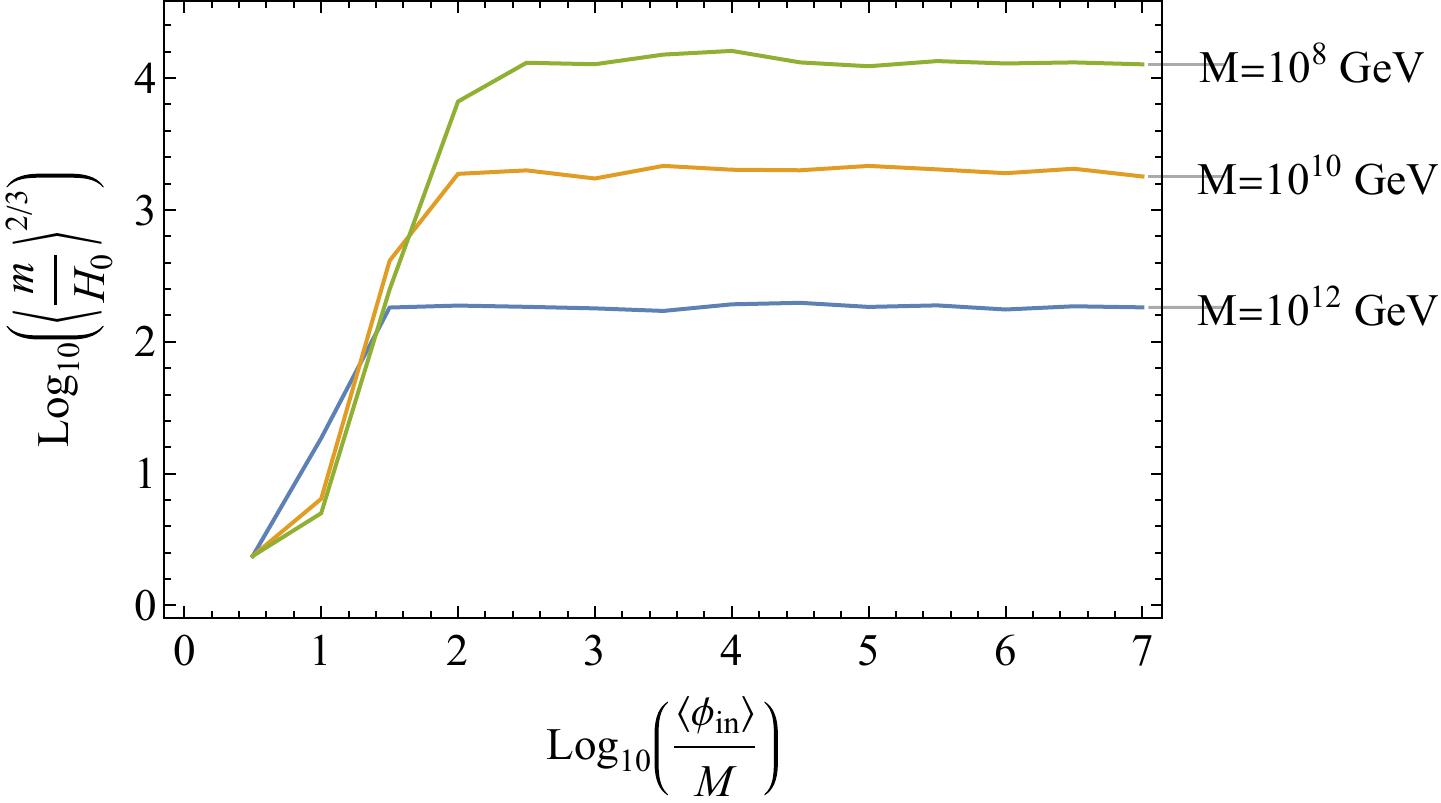}
    \caption{\small Average value of  $\left \langle (m/H_0)^{2/3}\right \rangle$ as a function of the mean initial field value $\left \langle\phi_{\rm in}/M\right \rangle$ for three selected values of the transition scale $M=10^{12}, 10^{10}, 10^8$ GeV. It is evident that lowering the transition scale increases the average value of the mass of the axions that become active after recombination.  } 
    \label{fig:parspacetrans}
\end{figure}

In direct analogy to Figure~\ref{fig:zerocorr} for the quadratic potential, we present the results for the number of axions and the corresponding abundance for the ``linear plus transition'' and ``linear'' cases.  We continue using the half-normal distribution for the initial field value and the log-normal distribution for the axion decay constant, as in Eq.~\eqref{eq:PDFs}, and introduce a random sign for each axion, to capture the effects of positive or negative initial field amplitude. As we discussed in the previous section, the bounds on the transition scale $M$ are different in the different regimes and so is the constraining region of $(\langle\phi_{\rm in}\rangle, \langle f_a\rangle)$. 
In Figures~\ref{fig:10Gev LPTres} and \ref{fig:10Gev linear} we fix $M=10^{10}$ GeV and show the results for the ``linear plus transition'' and ``linear'' case respectively. 
Note that the underlying contours for the numbers of axions in the $(\langle\phi_{\rm in}\rangle/M, \langle f_a\rangle/M)$ plane are the same for both cases since we approximate $\Delta\phi\simeq -\phi_{\rm in}$, but the constraints from the abundance change its final value  in the two cases. This is also clear from the different abundance contours of the right panels of  Figures~\ref{fig:10Gev LPTres} and \ref{fig:10Gev linear}. We can understand the slope of the constant $\Omega_\phi$-lines from Eqs.~\eqref{eq:ABlinear} and \eqref{eq:Ablinplustrans} through substituting the scaling 
$N\sim \left (\langle f_a\rangle / \langle \phi_{\rm in}\rangle \right )^2$.
The ``linear plus transition'' case is similar to the quadratic one, since the final abundance does not depend on the mass, thus we arrive at the following scaling 
\beq\label{eq:ABmeanLPT}
\langle\Omega_\phi\rangle\sim \frac{M^2}{M_{\rm Pl}^2}\left \langle
\frac{f_{a}^{2}}{M^{2}}
\right \rangle
\frac
{
\left \langle
\frac{\phi_{\rm in}^{15/6}}{M^{15/6}}
\right \rangle
}
{
\left \langle
\frac{\phi_{\rm in}^{2}}{M^{2}}
\right \rangle
}
 \, .
\eeq
Since the last factor of Eq.\eqref{eq:ABmeanLPT} grows with $\phi_{\rm in}$, equal $\langle\Omega_\phi\rangle$-lines show anti-correlation between $\langle\phi_{\rm in}\rangle$ and $\langle f_a\rangle$, as  can be seen in Figure~\ref{fig:10Gev LPTres}. On the other hand, the ``linear'' case is more complicated, because the final abundance, which scales as 
\beq\label{eq:ABmeanLin}
\langle\Omega_\phi\rangle\sim \frac{M^2}{M_{\rm Pl}^2}\left \langle
\frac{f_{a}^{2}}{M^{2}}
\right \rangle
\left \langle
\frac{
m^{2/3}}{H_0^{2/3}}
\right \rangle
\frac
{
\left \langle
\frac{\phi_{\rm in}^{5/3}}{M^{5/3}}
\right \rangle
}
{
\left \langle
\frac{\phi_{\rm in}^{2}}{M^{2}}
\right \rangle
}
 \, ,
\eeq
depends on the axion mass, whose mean value varies with the initial condition according to Eq.~\eqref{eq:tauosc1}.
In Figure~\ref{fig:parspacetrans} we show the average value $\left \langle (m/H_0)^{2/3}\right \rangle$ as a function of the initial condition\footnote{We compute the mean value of the mass for corresponding initial value by constructing many realizations of the pairs $(m/H_0,\phi_{\rm in}/M)$ taken from the green region of the allowed parameter space shown in Figure~\ref{fig:Backevol}. For each $\langle \phi_{\rm in}/M \rangle$ we select those values of $m/H_0$ within the $3\sigma_\phi$ band, and  construct a histogram of the corresponding masses that we then use for computing $\left \langle (m/H_0)^{2/3}\right \rangle$.  } for three different values of the transition scale $M$. Note that the mean mass increases for decreasing $M$, because higher values of $\phi_{\rm in}$ are allowed (without exceeding $\Omega_{\phi,{\rm max}}$) therefore more massive fields can contribute. Because of this dependence of the average mass on the initial condition, the final abundance shows a change of behaviour that shows up as a ``knee'' in Figure \ref{fig:10Gev linear}.
After this ``knee'' the average value of the mass grows slower with $\langle\phi_{\rm in}\rangle/M$ and the behaviour of the abundance is determined by the \jo{last factor} in Eq.~\eqref{eq:ABmeanLin}, which decreases with $\langle \phi_{\rm in}\rangle/M$. Then $\Omega_\phi$-lines show a positive correlation between $\langle\phi_{\rm in}\rangle$ and $\langle f_a \rangle$, as opposed to the case with the transition. Therefore, it is important to understand how different field evolution affects the degeneracy lines in the $(\langle \phi_{\rm in}\rangle,\langle f_a\rangle)$ parameter space and modifies the relative constraints. 
\begin{figure}
    \centering
    \begin{minipage}[b]{0.45\linewidth}
    \includegraphics[width=7.5cm]{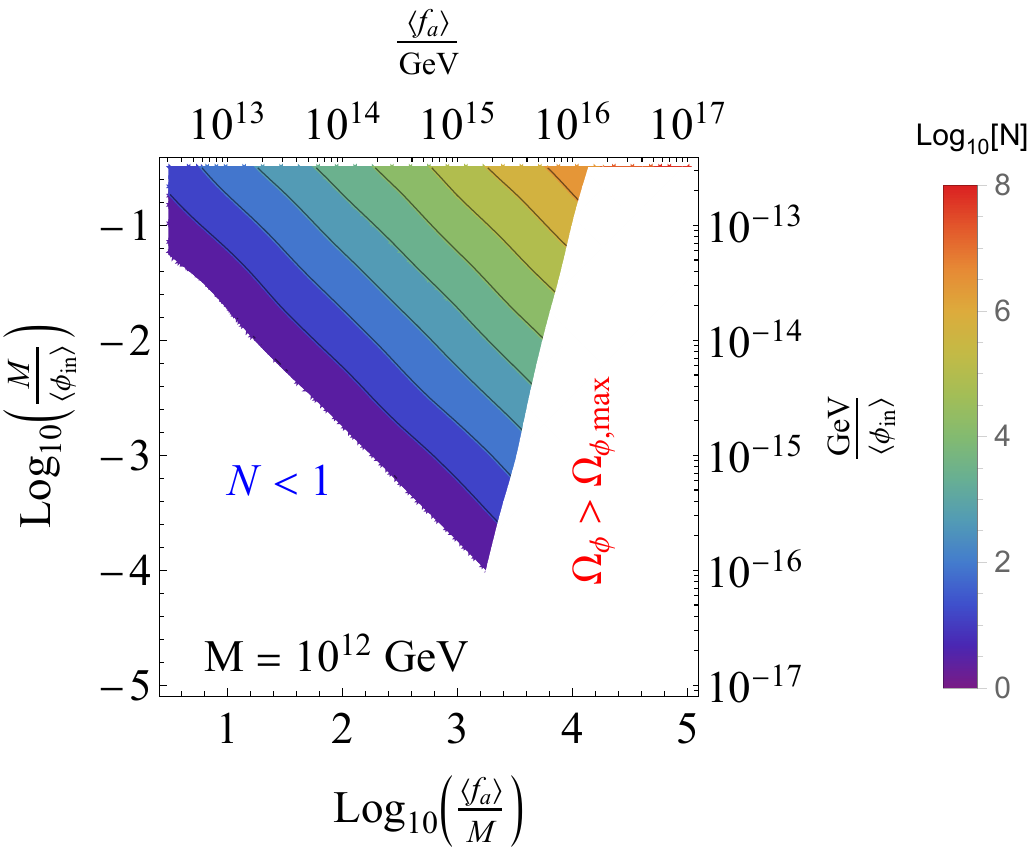}
    \end{minipage}
    \quad
    \begin{minipage}[b]{0.45\linewidth}
    \includegraphics[width=7.5cm]{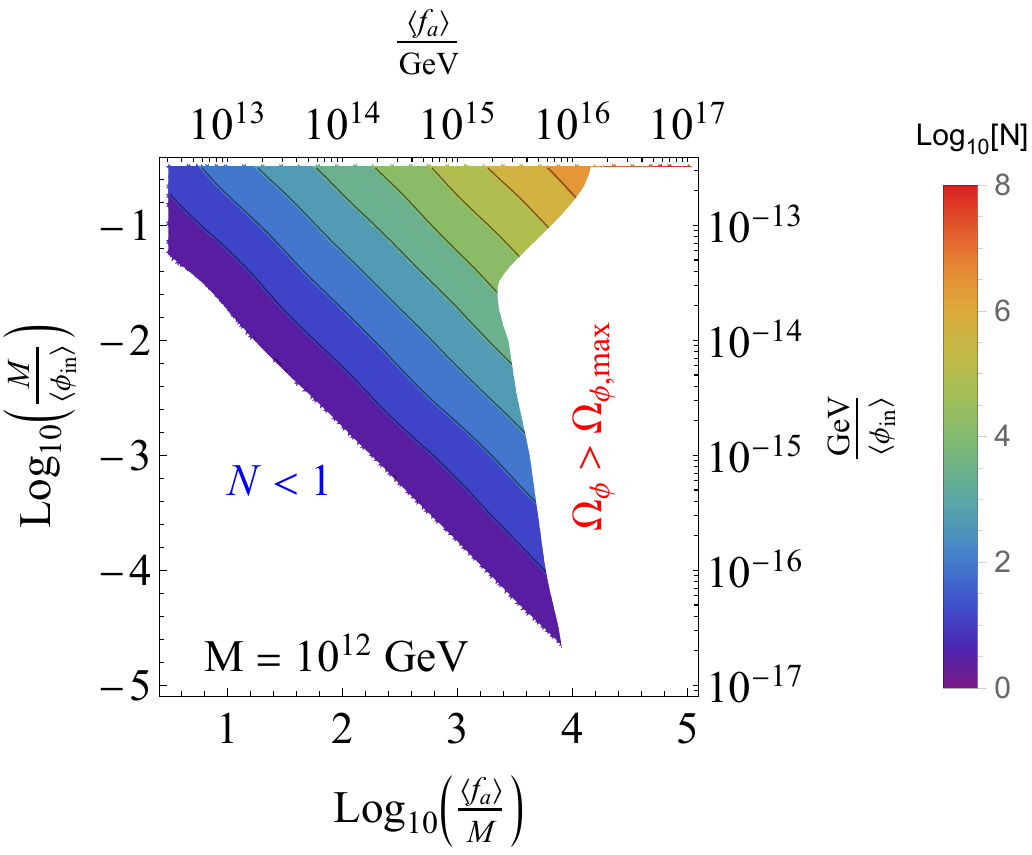}
    \end{minipage}
    \caption{\small The analogous plots to the left panels of Figures \ref{fig:10Gev LPTres} (left)  and \ref{fig:10Gev linear} (right)  of the ``linear plus transition'' and ``linear'' regime for a transition scale of $M=10^{12}$ GeV. }
       \label{fig:12Gevres}
\end{figure}

We now examine the dependence \jo{of our results} on the  transition scale $M$. Note that the dynamics of the scalar field, discussed in Section~\ref{sec:monodromy1/2}, does not depend on its absolute value, but on the relative values of $\langle \phi_{\rm in}\rangle/M$ and $\langle m\rangle/H_0$. The same holds for the birefringence angle, which depends on the relative values of $\langle\phi_{\rm in}\rangle/M$ and $\langle f_a\rangle/M$. 
The actual value of $M$ appears only in the  value of $\langle \Omega_\phi\rangle$. However, this turns out to be just a mild dependence.
For comparison, we present the above results for a different transition scale $M=10^{12}$ GeV in the left and right panels of Figure~\ref{fig:12Gevres} for the ``linear plus transition'' and ``linear'' case.
For the case with transition, by writing the initial value in terms of the Planck scale and inverting the Eq.~\eqref{eq:ABmeanLPT} for $\langle \Omega_\phi\rangle$, we find that the upper bound on the decay constant scales as \jo{$\langle f_{a}\rangle/M_{\rm Pl}\lesssim(M/M_{\rm Pl})^{1/4}$}, thus it decreases with the transition scale $M$. This expectation is confirmed by examining the tip of the allowed region in Figures~\ref{fig:12Gevres} and \ref{fig:10Gev LPTres}, which corresponds to $\langle f_{a}\rangle\simeq8\times10^{14}$ GeV for $M=10^{10}$ GeV and $\langle f_{a}\rangle\simeq2\times10^{15}$ GeV for $M=10^{12}$ GeV. We also note that the upper bound on the decay constant decreases with the transition scale for the linear case, where the analysis is more involved, because of the mass dependence. In this case, we obtain $\langle f_{a}\rangle\simeq4\times10^{15}$ GeV for $M=10^{10}$ GeV and $\langle f_{a}\rangle\simeq7\times10^{15}$ GeV for $M=10^{12}$ GeV. The bounds on the maximum initial field value and the corresponding decay constant are summarized in Table \ref{tab:mon}. We see that changing the transition scale $M$ by two orders of magnitude,  mildly affects the other parameters. 
\\ 

\begin{table}
\centering
\begin{tabular}{|p{5cm}|p{5cm}|p{5cm}|}
\hline

          & \textbf{M$\mathbf{=10^{10}}$ GeV} &  \textbf{M$\mathbf{=10^{12}}$ GeV}\\
  \hline\hline
  \textbf{Linear plus transition}
  & $\langle f_{a}\rangle\simeq8\times10^{14} {\rm GeV}$ & $\langle f_{a}\rangle\simeq2\times10^{15} {\rm GeV}$\\
  & $\langle\phi_{\rm in, max}\rangle<8\times10^{15} {\rm GeV}$  &  $\langle\phi_{\rm in, max}\rangle<10^{16} {\rm GeV}$ \\
                          
  \hline
  \textbf{Linear}      
  & $\langle f_{a}\rangle\simeq4\times10^{15} {\rm GeV}$ & $\langle f_{a}\rangle\simeq7\times10^{15} {\rm GeV}$  \\
  & $\langle\phi_{\rm in, max}\rangle<7\times10^{16} {\rm GeV}$  &  $\langle\phi_{\rm in, max}\rangle<8\times10^{16} {\rm GeV}$ \\
                         
  \hline
\end{tabular}
\caption{\small Summary of the maximum values for the initial field value and corresponding decay constant for the ``linear plus transition'' and ``linear'' regime discussed in section \ref{sec:monodromy1/2}. These values correspond to the \jo{bottom corner } of the allowed regions shown in Figures~\ref{fig:10Gev LPTres}, \ref{fig:10Gev linear} and~\ref{fig:12Gevres}.}\label{tab:mon}
\end{table}

\section{Summary and discussion}
\label{sec:summary}

In this work, we explore the expectations and the implications for the multidimensional probability density functions of the axions' parameters as the initial field value, the decay constant and the mass in the Axiverse scenario. We consider three different types of axion potential:  cosine,  quadratic  and  asymptotically linear monodromy potential.
In most cases, the average value of the birefringence angle vanishes, since axion contributions with  
different sign cancel each other out on average. Therefore, except in special ``aligned'' cases, we consider the standard deviation of the distribution of values of the birefringence angle as the characteristic value.  

In the case of the usual cosine potential, we  note that more than $\mathcal{O}(20)$ axions are needed to explain the observed value of the birefringence angle, thus a statistical treatment is justified. In the multidimensional parameter space resulting from string compactifications, correlations might appear and we studied the effect of those on the number of inferred axions. In the Gaussian approximation, we give analytical formulas for the variance of the birefringence angle with a correlation between the initial misalignment and the anomaly coefficient that can change the dependence on the number of axions from $\sqrt{N}$ to linear in $N$, where $N$ is the number of axions that contribute to the birefringence signal. 
We also showed how to use future observational data, providing the ratio between birefringence from reionization and recombination, to probe the distribution of axion masses, \jo{given a model for the axion potential and couplings.} In the case of a log-uniform distribution, there is a clear expectation of $\beta_{\rm rei}/\beta_{\rm rec}\simeq0.7$.  Deviations from this result can be parameterized using a tilt in the mass distribution, which can be compared to expectations in the context of the string  Axiverse.

The case of a quadratic potential allows for significant analytical treatment since  the axion evolution can be derived analytically and thus we can connect the final abundance to the initial field amplitude. Limits on the present abundance provide an effective maximum displacement since each field contribution adds up linearly as opposed to the variance of the birefringence. Because of the different dependence on $N$, the abundance provides an upper bound on the value of the decay constant. \jo{If we take the decay constant to be the same for all axions, (defined in Eq.~\eqref{eq:lagrangian}) } we derive the limit $f_a\lesssim 2\times 10^{16}$ GeV in the uncorrelated scenario, \jo{ which}  is modified by an $\mathcal{O}(1)$ coefficient by introducing correlations. This is stricter than the corresponding constraint derived in the single axion case by an order of magnitude.
Interestingly, we found that for large (allowed) values of the decay constant $f_a\sim10^{16}$ GeV, there must be an upper limit of the mass distribution around $m=10^{-24}$ eV, coming from the expectation value of the projected abundance. 
This is an interesting model-building tool, as it strongly ties the axion decay constant to the mass distribution of axions.
This maximum allowed mass increases as we decrease $f_a$. Furthermore, we showed how the correlations between the axion masses and the decay constant or the initial displacement affect the birefringence tomography ratio  $\beta_{\rm rei}/\beta_{\rm rec}$.  When the sign of the initial field amplitude is correlated with the axion mass, the birefringence ratio can even become negative. This is an important way to support or rule out certain axiverse scenarios, using future observational data.

The case of the axion monodromy potential is more complicated, since one needs to take into account one extra mass scale, the potential turnover scale, and the full scalar field evolution is not analytically known. 
In particular, the evolution  depends  on the relative values of the initial field and the  potential turnover scale. We found the region of the parameter space in which the fields evolve solely in the linear or quadratic regime of the potential and when there is a transition of behaviour. We derived analytical estimates for the final abundance in each case and  determined  the available region to explain the birefringence evidence. As an example of the difference between the axion monodromy and quadratic potential, the axion mass for the former can be two orders of magnitude larger than that of the latter, depending on the initial field amplitude. This significantly affects the complementarity between cosmological and laboratory axion searches.

We demonstrate that, although the actual shape of the contours depends on the model, an upper bound of $f_a\lesssim \mathcal{O}(1)10^{16}$ GeV seems to remain in all cases and gets mildly stronger (lower) for smaller values of the transition scale $M$, which controls the turnover between the quadratic and linear parts of the potential.   {Before concluding, we should comment on the possible breakup of the axion condensates due to parametric resonance in anharmonic potentials, like axion monodromy \cite{Amin:2011hj}. A possible breakup of the condensate can lead to GW signals, however this may require a flat potential with a negative second derivative for some range of field values \cite{Kitajima:2018zco}. Even in this case, the expected GW frequency will be too small to be detectable, since the process would take place during matter domination. Furthermore, such a break-up of the axion condensate would not affect our conclusions, since birefringence depends on the rolling of the axion from its initial amplitude until today or until the minimum of the potential, whatever occurs first.}

This work represents an initial attempt of using the cosmic birefringence data to constrain the properties of the Axiverse.
In order to keep the analysis tractable, we limited ourselves to specific potentials and considered a limited number of input random variables in each case. Despite the expectation of a universe filled with (pseudo)scalar fields, most analyses of axion phenomenology have focused either on single or few-field models (with important exceptions, e.g. Refs.~\cite{Mehta:2020kwu, Mehta:2021pwf}, focusing on black hole superradiance). On the contrary, multi-field dynamics has been extensively applied to inflationary model-building \cite{Dias:2017gva, Aragam:2019omo, Paban:2018ole, Christodoulidis:2019hhq, Bachlechner:2017hsj, Bachlechner:2017zpb}, thus providing interesting parallels between the two areas of physics. A similar research program is necessary, in order to fully appreciate the implications of an axiverse (whether of string theory origin or not). \jo{Recently, machine learning methods have found applications in physics~\cite{He:2023csq}, specifically in explorations of the String landscape~\cite{Carifio:2017bov, He:2022cpz}, as well as in reconstructing the inflationary potential from cosmological data~\cite{Kamerkar:2022dfu}. 
Building on these advances presents an interesting avenue for a more extensive evaluation of the multi-dimensional model space that Axiverse models possess, including  axions with different potentials and correlations between several of the parameters.
}

\section*{Acknowledgements}
 We are indebted to E.~Komatsu for detailed comments on the manuscript. We  are also grateful to I.~Obata for
suggesting a more in-depth analysis of the birefringence tomography and to A. Maleknejad for the interesting discussion on the Cosmic Birefringence as a ``memory'' effect.
We would also like to thank D.~Blas, R.~Ferreira and O.~Pujolas  for useful discussions and comments of the manuscript. 
\jo{Finally, we thank the anonymous referee for providing constructive feedback on the manuscript.}
EIS is supported by a fellowship from ``la Caixa'' Foundation (ID 100010434) and from the European Union’s Horizon 2020 research
and innovation programme under the Marie Sk\l{}odowska-Curie grant agreement No 847648, the fellowship code is
LCF/BQ/PI20/11760021. SG has the support of the predoctoral program AGAUR FI SDUR 2022 of the Secretariat of Universities and Research of the Department of Research and Universities of the Generalitat of Catalonia and the European Social Plus Fund.

\appendix

\section{ Propagation of variance and correlations}
\label{app:corralation}

For completeness and notational clarity, we define the basic statistical concepts that are used throughout the paper. 

Consider a random variable $x$ with a probability density function $f(x)$, defined over a finite or infinite domain $A$, such that $\int_A f(x)\,dx=1$.
The expectation value, or average or mean, is defined as
$\langle x \rangle \equiv E(x) =\int_A x\cdot f(x)\, dx $, while the variance is 
$ {\rm Var}(x) = \langle x^2\rangle - \langle x\rangle^2$ and the standard deviation is $\sigma_x \equiv \sqrt {{\rm Var}(x) }$.

We can estimate the variance of a random variable $y=g(x)$ by using a Taylor expansion around the mean of $x$
\begin{equation}
    g(x)=g(\langle x\rangle)+g'(\langle x \rangle)\left ( x-\langle x\rangle \right )+\frac{1}{2}g''(\langle x \rangle)\left(x-\langle x\rangle\right )^2+...
\end{equation}
By keeping only the linear term, we can approximate the average as $\langle g (x)\rangle \simeq g\left (\langle x \rangle \right )$ and the
variance as
\beq
{\rm Var}[y]\equiv \langle y^2\rangle - \langle y\rangle^2 
\simeq \left [g'(\langle x\rangle) \right ]^2 \left (
\langle x^2\rangle - \langle x\rangle^2
\right ) = \left [g'(\langle x\rangle) \right ]^2 {\rm Var}[x] \, .
\eeq
For a random variable $y$, which is itself a function of two random variables $x_1$ and $x_2$,
$y=g(x_1,x_2)$, the PDF is Taylor-expanded as
\beq
g(x_1,x_2)=\langle g\rangle +g_{x_1}(x_1-\langle x_1\rangle)+g_{x_2}(x_2-\langle x_2\rangle)+g_{x_1, x_2}(x_1-\langle x_1\rangle)(x_2-\langle x_2\rangle)+...
\eeq
where we define $ \langle g\rangle \equiv g(\langle x_1\rangle,\langle x_2\rangle)$. Thus its expectation value reads
\beq
{\rm E}[g]=\langle g\rangle +g_{x_1, x_2}{\rm Cov}[x_1,x_2]
\eeq
and the variance is
\beq
\begin{split}\label{eq:expvar}
    {\rm Var}[g]& \simeq g^2_{x_1}{\rm Var}[x_1]+g^2_{x_2}{\rm Var}[x_2]+2{\rm Cov}[x_1,x_2]g_{x_1}g_{x_2}+\\
    &+2g_{x_1, x_2}(\langle x_1^2x_2\rangle g_{x_1}+\langle x_1x_2^2\rangle g_{x_2})+({\rm Cov}[x^2_1,x^2_2]-{\rm Cov}[x_1,x_2]^2+{\rm Var}[x_1]{\rm Var}[x_2])g^2_{x_1,x_2},
\end{split}
\eeq
where terms in the first line are at first order and in the second line at second.
The covariance is defined as
\beq
\text{Cov}(x_1,x_2) \equiv \langle
(x_1-\langle x_1 \rangle ) (x_2-\langle x_2\rangle)
\rangle
\, .
\eeq
and the correlation between two random variables is 
\beq
\rho(x_1,x_2) \equiv\frac{\text{Cov}(x_1,x_2)}{\sigma_{x_1}\sigma_{x_2}}
\eeq

Specifically in the case of the product of two Gaussian random variables the mean and the variance can be 
computed exactly, without resorting to the Taylor expansion approximation. This is because Gaussian PDFs are uniquely defined solely in terms of their mean and variance, without the need for higher order correlators.
Assume two random Gaussian random variables $x,y$, with mean $X,Y$ and variances $\sigma_x^2, \sigma_y^2$. The mean and variance of the product $xy$ are:
\begin{align}\label{eq:A89}
    &{\rm E}[xy]= XY+ \rho\sigma_x\sigma_y\\
    &\sigma_{xy}^2=X^2\sigma_y^2+Y^2\sigma_x^2 +2XY\rho\sigma_x\sigma_y+\sigma_x^2\sigma_y^2(1+\rho^2).
\end{align}
If the mean values of $x$ and $y$ are zero then we have ${\rm E}[xy]= \rho\sigma_x\sigma_y$ and $\sigma_{xy}=\sigma_x\sigma_y\sqrt{(1+\rho^2)}$. In the text, we introduced the sign of an additional random variable that is independent of $(x,y)$ that we multiply such that ${\rm sgn}(z)xy$. In this case, the mean of the product becomes zero even for a non-zero correlation of $(x,y)$, but the variance gets an extra contribution, as  can be seen from \jo{Eq.~\eqref{eq:A89}} with $x_1=z$ and $x_2=xy$:
\beq\label{eq:varwithsgn}
{\rm Var}[x_1 x_2]= {\rm E}^2(x_2)+{\rm Var}(x_2)\, ,
\eeq
where we used the fact that the mean value of the sign is zero and its variance is one. 
\begin{figure}
    \centering
    \includegraphics[scale=0.7]{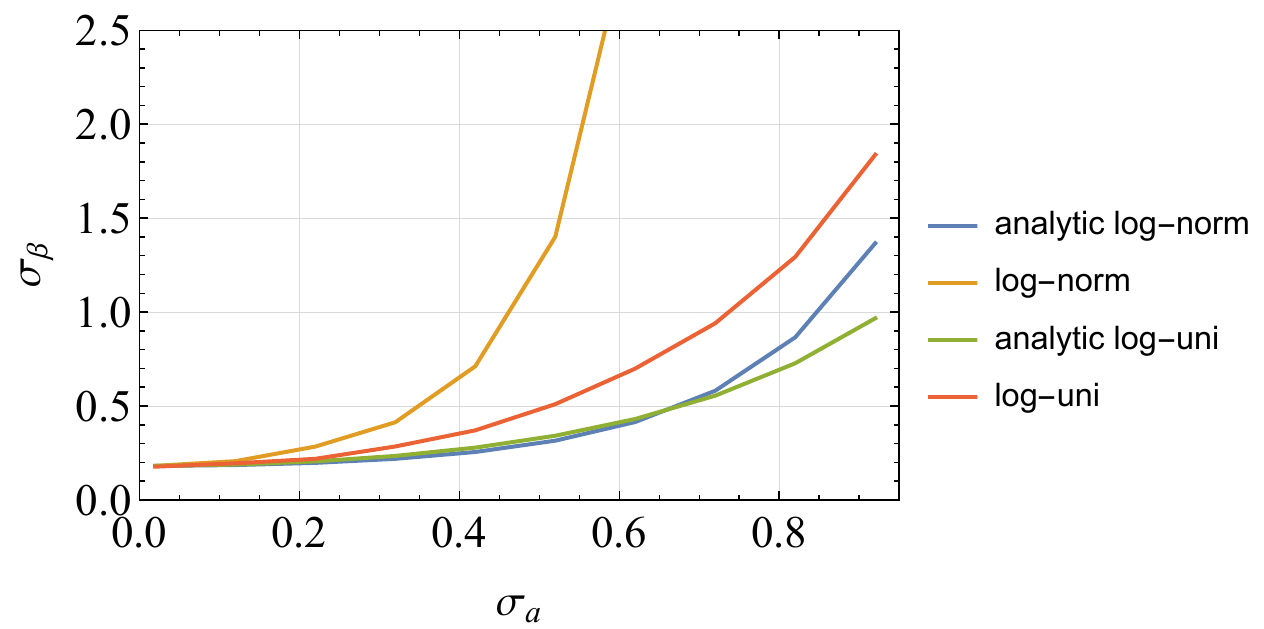}
    \caption{\small We show the standard deviation  $\sigma_\beta$  of the birefringence PDF $\sigma_\beta$ as a function of $\sigma_{a}$ for the log-normal and the log-uniform distribution and  the comparison with the analytical formulas given in Eq.~\eqref{eq:analyticvar}. The parameters of the distributions are those of Figure~\ref{fig:birrpdf}.
    }
    \label{fig:comparison Sigma}
\end{figure}
\jo{In Figure\ref{fig:comparison Sigma} we show the comparison between the numerical computation of the variance of $\beta$ and the analytical formula derived by \eqref{eq:A89} and \eqref{eq:varwithsgn} applied to the case of interest which gives \eqref{eq:analyticvar}. We also compare the result of taking a uniform or normal distribution in logarithmic space for the decay constant, for the normal case the numerical $\sigma_\beta$ starts diverging quicker from the analytic formula with the increase of $\sigma_a$ because for long tails the expansion used \eqref{eq:expvar} is not accurate. }

\section{Constructing joint distributions}
\label{app:copulas}

\begin{figure}
   \begin{tabular}{ccc}
\subfloat[$\rho=0$]{\includegraphics[width = 2in]{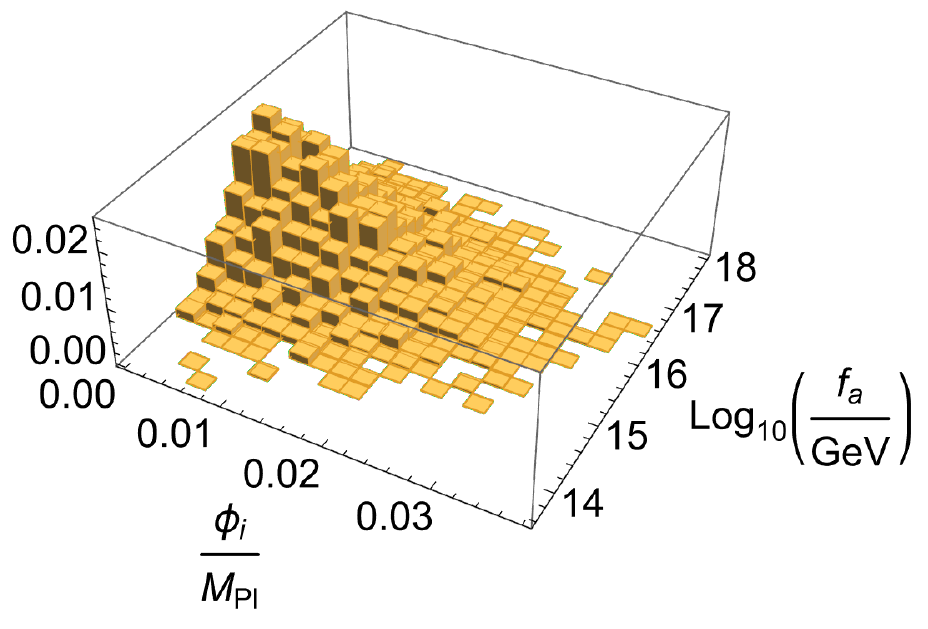}} &
\subfloat[$\rho=0.9$]{\includegraphics[width = 2in]{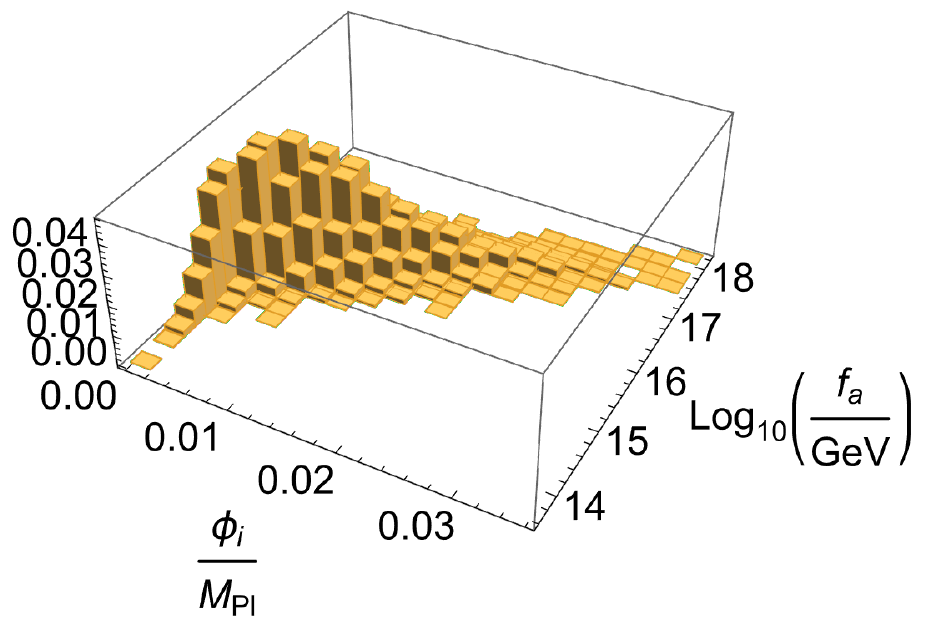}} &
\subfloat[$\rho=-0.9$]{\includegraphics[width = 2in]{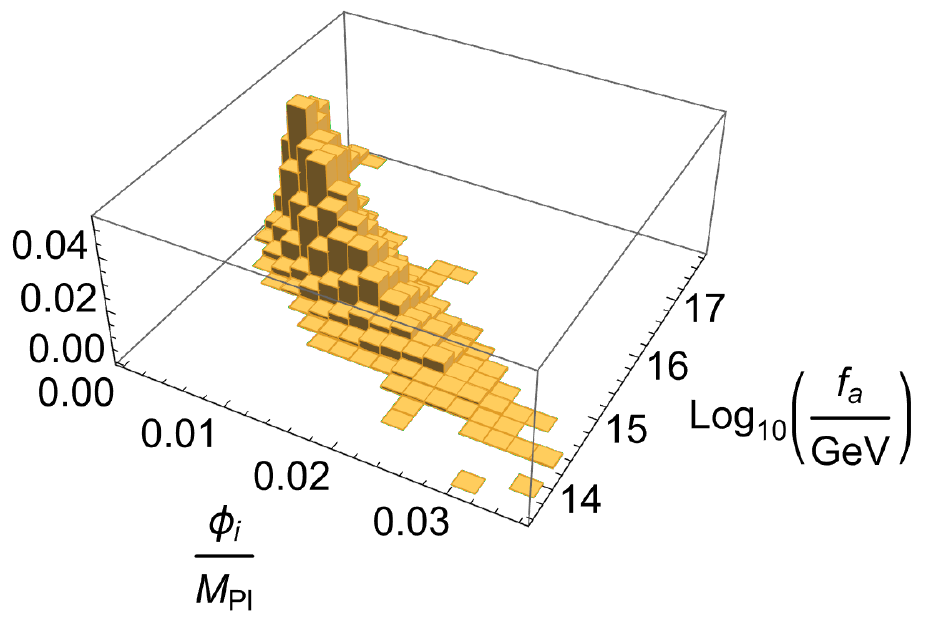}}
\end{tabular}
\caption{Joint PDF of the decay constant, whose marginal is a Gaussian distribution in log-space, and initial condition, whose marginal is a half Gaussian distribution. The marginal distribution of the decay constant is taken with $\log_{10}{(\langle f_a\rangle/\text{GeV})}=16$ and dispersion $\sigma_{a}=0.25$, for the initial field value $\sigma_{\phi_i}=10^{-2}$. In the second and third panels from the left, we can clearly see the difference between choosing a positive $(\rho=0.9)$ or a negative correlation  $(\rho=-0.9)$ between the two variables.
}
\label{fig:PDFVisual}
\end{figure}

To introduce correlations between the different random variables, characterized by their marginal distributions, we use the copula built-in function in {\it Mathematica}. The idea of copulas in statistics is related to Sklar's theorem~\cite{sklar}, which argues that for any multivariate distribution of $d-$random variables $\Hat{f}_{X_1,...X_d}$, characterized by marginal PDFs $\Hat{f}_{X_1}(x_1)...\Hat{f}_{X_d}(x_d)$, there exists a function $\mathcal{C}$ such that 
\begin{equation}
   \Hat{f}_{X_1,...X_d}=\mathcal{C}(\Hat{f}_{X_1}(x_1)...\Hat{f}_{X_d}(x_d)). 
\end{equation}
Formally, the copula $\mathcal{C}(u_1,..u_d)$ is any joint distribution whose marginals are uniform in $[0,1]$, i.e. $\mathcal{C}: (u_1,..u_d)\in [0,1]^d \rightarrow [0,1]$ and it can be used for an arbitrary marginal distribution $\Hat{f}_{X}$ by mapping its input parameters to those of the copula via the cumulative density function $\pi_X=\int_{x_{min}}^x{\rm d}x' \Hat{f}_{X}(x')$. In applied statistics, several copulas exist which encode different degrees of dependence between variables. In this work we use the Gaussian copula, a frequently used one with no particular weighting of the different parts of the distribution. Formally, for two random variables, this is defined as 
\begin{equation}
    \mathcal{C}(u_1,u_2,\rho)=\Phi_\rho(\Phi^{-1}(u_1),\Phi^{-1}(u_2)), 
\end{equation}
where $\Phi_\rho$ is the standard bivariate normal distribution with correlation $\rho$ and $\Phi^{-1}$ is the inverse of the simple univariate normal distribution function. Figure~\ref{fig:PDFVisual} shows some examples of a joint PDF, using a Gaussian copula, with different values of the correlation parameter.

\section{Numerical details}
\label{app:numdet}
The numerical procedure used to produce the contours and the exclusion plots for $(\phi_{\text{in}},f_a)$ as Figures \ref{fig:zerocorr}, \ref{fig:poscorr} and \ref{fig:negcorr}, \jo{in the non-``aligned'' case}, is the following:
\begin{itemize}
    \item Starting from the \jo{marginal} distributions on the initial field value and the decay constant given in Eq.\eqref{eq:PDFs} we construct a 2D grid by varying the input parameters of the distribution which are related to the average values as $\langle  \phi_{\text{in}}\rangle =\mu_\phi$ and $\langle f_a\rangle=\exp{(\mu_{a}+\sigma_a^2/2)}$. We vary them as $\mu_\phi=10^{-m}$, in Planck units, with $m\in[0.3,4.5]$  and $\mu_{a}=\ell/\log_{10}(e)-(0.25/\log_{10}(e))^2/2$ and vary $\ell \in [13.5,17]$, units are in GeV\footnote{The factor $\log_{10}(e)$ is because $\ell$ is in base 10, thus also the input deviation parameter is $\sigma/\log_{10}(e)$}. Starting from these marginal distributions we build the joint distribution with correlation via the built-in binormal copula distribution function in {\tt Mathematica}. We briefly review the concept of copula distribution in Appendix \ref{app:copulas}.  
    
    \item The birefringence is computed using Eq.~\eqref{eq:betabidis} for $N$  axions over several realizations.  For each point of the grid, we then compute the value of $\sigma_\beta^2\propto N$ from interpolating the variance of the resulting distribution of $\beta$ for an increasing number of axions.

    \item We find the number of axions needed for each parameter combination (grid point) by requiring that $\sigma_\beta(\mu_a,\mu_\phi,N)=0.3$ deg.  We then construct the contour plot coming from the 2D interpolation of $N(\mu_a,\mu_\phi,\beta_{\rm obs})$ over the grid.

    \item Concerning the abundance, we interpolate the mean value of $\Omega_\phi$, for each $\mu_\phi$, as a function of an increasing number of axions. Subsequently, we compute the corresponding abundance for each grid point by evaluating $\Omega_\phi(\mu_\phi,\mu_a, N)$. 
\end{itemize}
\jo{In the ``aligned'' case we take $\beta\sim \langle \beta\rangle\pm\sigma_\beta $ and instead of using the fit $\beta\sim N$, we find the parameters of the more complicated fit 
\begin{equation}\label{eq:Nscalealign}
\beta=a N+b\sqrt{N}.
\end{equation}
We confirmed that this model recovers well the scaling of $\beta$ in this case. The following steps are the same as previously described with N coming from inverting $\beta=0.3$ deg, together with Eq.~\eqref{eq:Nscalealign}. }

\bibliography{stringbirefringence}

\end{document}